\documentclass[sigconf,screen]{acmart}
\usepackage{algorithmic}
\usepackage{graphicx}
\usepackage{textcomp}
\usepackage{xcolor}
\usepackage[inline,shortlabels]{enumitem}
\usepackage{tcolorbox}
\usepackage{booktabs}
\usepackage{graphicx}
\usepackage[table]{xcolor} %
\usepackage{pifont}
\usepackage{pgfplots}
\usetikzlibrary{patterns} 
\pgfplotsset{compat=1.17}
\usepackage{subcaption}
\usepackage{enumerate}
\usepackage{svg}
\usepackage{url}
\usepackage{microtype}
\usepackage[capitalise]{cleveref}

\pgfplotsset{
    box plot/.style={
        /pgfplots/.cd,
        black,
        only marks,
        mark=-,
        mark size=\pgfkeysvalueof{/pgfplots/box plot width},
        /pgfplots/error bars/y dir=plus,
        /pgfplots/error bars/y explicit,
        /pgfplots/table/x index=\pgfkeysvalueof{/pgfplots/box plot x index},
    },
    box plot box/.style={
        /pgfplots/error bars/draw error bar/.code 2 args={%
            \draw  ##1 -- ++(\pgfkeysvalueof{/pgfplots/box plot width},0pt) |- ##2 -- ++(-\pgfkeysvalueof{/pgfplots/box plot width},0pt) |- ##1 -- cycle;
        },
        /pgfplots/table/.cd,
        y index=\pgfkeysvalueof{/pgfplots/box plot box top index},
        y error expr={
            \thisrowno{\pgfkeysvalueof{/pgfplots/box plot box bottom index}}
            - \thisrowno{\pgfkeysvalueof{/pgfplots/box plot box top index}}
        },
        /pgfplots/box plot
    },
    box plot top whisker/.style={
        /pgfplots/error bars/draw error bar/.code 2 args={%
            \pgfkeysgetvalue{/pgfplots/error bars/error mark}%
            {\pgfplotserrorbarsmark}%
            \pgfkeysgetvalue{/pgfplots/error bars/error mark options}%
            {\pgfplotserrorbarsmarkopts}%
            \path ##1 -- ##2;
        },
        /pgfplots/table/.cd,
        y index=\pgfkeysvalueof{/pgfplots/box plot whisker top index},
        y error expr={
            \thisrowno{\pgfkeysvalueof{/pgfplots/box plot box top index}}
            - \thisrowno{\pgfkeysvalueof{/pgfplots/box plot whisker top index}}
        },
        /pgfplots/box plot
    },
    box plot bottom whisker/.style={
        /pgfplots/error bars/draw error bar/.code 2 args={%
            \pgfkeysgetvalue{/pgfplots/error bars/error mark}%
            {\pgfplotserrorbarsmark}%
            \pgfkeysgetvalue{/pgfplots/error bars/error mark options}%
            {\pgfplotserrorbarsmarkopts}%
            \path ##1 -- ##2;
        },
        /pgfplots/table/.cd,
        y index=\pgfkeysvalueof{/pgfplots/box plot whisker bottom index},
        y error expr={
            \thisrowno{\pgfkeysvalueof{/pgfplots/box plot box bottom index}}
            - \thisrowno{\pgfkeysvalueof{/pgfplots/box plot whisker bottom index}}
        },
        /pgfplots/box plot
    },
    box plot median/.style={
        /pgfplots/box plot,
        /pgfplots/table/y index=\pgfkeysvalueof{/pgfplots/box plot median index}
    },
    box plot width/.initial=1em,
    box plot x index/.initial=0,
    box plot median index/.initial=1,
    box plot box top index/.initial=2,
    box plot box bottom index/.initial=3,
    box plot whisker top index/.initial=4,
    box plot whisker bottom index/.initial=5,
}

\copyrightyear{2025}
\acmYear{2025}
\setcopyright{cc}
\setcctype{by}
\acmConference[IOT 2025]{The 15th International Conference on the Internet of Things}{November 18--21, 2025}{Vienna, Austria}
\acmBooktitle{The 15th International Conference on the Internet of Things (IOT 2025), November 18--21, 2025, Vienna, Austria}
\acmPrice{}
\acmDOI{10.1145/3770501.3770515}
\acmISBN{979-8-4007-1595-2/25/11}

\begin{document}

\title{Lumos: Performance Characterization of WebAssembly as a Serverless Runtime in the Edge-Cloud Continuum}

\author{Cynthia Marcelino}
\orcid{0000-0003-1707-3014}
\affiliation{%
  \department{Distributed Systems Group}
  \country{TU Wien, Vienna, Austria} 
  }
\email{c.marcelino@dsg.tuwien.ac.at}

\author{Noah Krennmair}
\orcid{0009-0001-2377-2826}
\affiliation{%
    \department{Distributed Systems Group}
    \country{TU Wien, Vienna, Austria} 
  }
\email{e12023994@student.tuwien.ac.at}

\author{Thomas Pusztai}
\orcid{0000-0001-9765-6310}
\affiliation{%
    \department{Distributed Systems Group}
    \country{TU Wien, Vienna, Austria} 
  }
\email{t.pusztai@dsg.tuwien.ac.at}

\author{Stefan Nastic}
\orcid{0000-0003-0410-6315}
\affiliation{%
    \department{Distributed Systems Group}
    \country{TU Wien, Vienna, Austria} 
  }
\email{snastic@dsg.tuwien.ac.at}

\begin{abstract}
WebAssembly has emerged as a lightweight and portable runtime to execute serverless functions, particularly in heterogeneous and resource-constrained environments such as the Edge–Cloud Continuum. However, the performance benefits versus trade-offs remain insufficiently understood. Current state-of-the-art performance characterizations either focus on isolated WebAssembly benchmarks or fail to address workload diversity, such as compute-intensive and data-intensive scenarios, in dynamic and heterogeneous environments such as the Edge-Cloud Continuum.
Therefore, in this paper, we present Lumos, a performance model and benchmarking tool for characterizing serverless runtimes. Lumos identifies workload, system, and environment-level performance drivers in the Edge-Cloud Continuum. We benchmark state-of-the-art containers and the Wasm runtime in interpreted mode and with ahead-of-time compilation. Our performance characterization shows that AoT-compiled Wasm images are up to 30x smaller and decrease cold-start latency by up to 16\% compared to containers, while interpreted Wasm suffers 30–55x higher warm latency and 7–10x I/O-serialization overhead. 
\end{abstract}


\maketitle

\newcommand{\cmark}{\ding{51}} 
\newcommand{\xmark}{\ding{55}} 

\renewcommand{\shortauthors}{Marcelino et al.}

\section{Introduction}

Serverless computing provides infrastructure management, elastic scaling, and on-demand resource allocation. In serverless computing, user code is deployed as small, short-lived functions that respond to events, scaling up or down according to demand~\cite{TheGapResearchRealWorld,serverless-why-when-how,whatServerlessIsAndWhatShouldBecome}. 
Recently, Wasm has emerged, offering serverless functions, strong isolation, and decreased cold starts. In Wasm-based functions, the code is compiled into small binary modules, which are executed within a lightweight, memory-safe, and secure sandbox, also known as the Wasm VM, enabling cross-platform portability while maintaining minimal overhead~\cite{WasmExecutionModel,PushingWasmEdge,wasmCommonLayer}. Moreover, Wasm follows a deny-by-default principle, which means Wasm VMs do not have access to the host unless explicitly allowed. To access host features such as network interfaces and filesystem, Wasm VMs leverage the WebAssembly System Interface (WASI)~\cite{CrossArchitectureWasmECC,faasm,sledge}.  

Wasm enables next-generation serverless computing by offering a portable, secure, and lightweight execution model, making it well-suited for heterogeneous, resource-constrained environments such as the Edge–Cloud Continuum
~\cite{wasmCommonLayer,WasmExecutionModel,Cwasi2023,goldfish2024}. However, its performance characteristics in the Edge-Cloud Continuum remain insufficiently understood.
Existing performance characterization of Wasm includes: 
\begin{enumerate*} [label=(\alph*)]
    \item \emph{Standalone Runtime Analysis}~\cite{wang_how_far_2022,WasmNextGen,jangda_not-so-fast_2019,BenchmarkingWasm2025,RevealingWasmPerformance} focus on low-level Wasm runtime behavior relative to native execution, typically measuring execution latency, instruction count, and CPU and memory usage. However, they do not effectively evaluate the performance of Wasm applications when deployed as OCI bundles, nor compare against Docker containers, which is the state-of-the-art runtime execution in the Edge-Cloud Continuum. Understanding how Wasm-based runtimes perform in this context is essential for evaluating the potential overhead in serverless platforms that rely on container-based isolation.
    \item \emph{Serverless Runtimes} ~\cite{WebAssemblyOrchestration,ContainerRuntimeYet2025,CrossArchitectureWasmECC,EvaluatingWasmEdge} measure cold start latency, execution time, and warm start performance of functions deployed in serverless environments. However, they do not consider the impact of I/O and (de)serialization overhead, particularly in Wasm runtimes that use WASI for networking and file access. Thus, it limits a detailed identification of performance bottlenecks in serverless functions, which typically rely on remote services for data exchange.
\end{enumerate*}

While Wasm promises near-native speed, it is essential to understand the benefits and tradeoffs associated with it in order to build smart systems to optimize performance. Therefore, in this paper, we propose Lumos, a performance model and framework that enables performance characterization of serverless functions, leveraging multiple serverless runtimes. Lumos enables users to evaluate serverless functions with different workload characteristics, such as compute and data-intensive, as well as measure data IO and serialization, which incurs a great impact on the performance of serverless functions in the Edge-Cloud Continuum~\cite{Truffle2024,sand,faastlane}. 

We summarize our contributions as follows:
\begin{itemize}
    \item \emph{Lumos: A Performance Model} for serverless functions in the Edge–Cloud Continuum. We identify primary performance factors across workload, system, and environment. The Lumos model captures the influence of these dimensions on Service Level Objectives (SLOs) such as latency, throughput, and cost. Thus, enabling the identification of performance bottlenecks in the Edge-Cloud Continuum.
    \item \emph{A benchmarking tool for evaluating serverless runtimes}. Lumos benchmark tool supports reproducible experiments across cold and warm starts, diverse workloads such as compute and data-intensive, and varying concurrency levels in heterogeneous environments;
    \item \emph{A performance characterization of WebAssembly as serverless runtime}, examining their execution models, cold starts, isolation mechanisms, and integration within the serverless platforms. 
    Our results show that Wasm Ahead-of-Time (AoT) images are up to 30x smaller and decrease cold-start latency by up to 16\% compared to containers, while
    Wasm interpreted suffers 30–55x higher warm latency and 7–10x I/O-serialization overhead.
\end{itemize}

\section{Research Challenges}\label{sec:background}

In this section, we identify the following research challenges of Serverless function performance in the Edge-Cloud Continuum.


\paragraph{RC-1: How do workload, system, and environmental factors influence the performance of serverless functions across the Edge–Cloud Continuum?}

Serverless function performance typically relies on Service Level Objectives (SLOs) such as end-to-end execution time, cost, and throughput to evaluate its performance~\cite{nastic2020sloc}. However, SLOs are affected by various interdependent factors that range from business logic to the underlying systems and environment. For example, workload characteristics such as data IO can impact function execution time by up to 95\%~\cite{sand,faastlane}. Additionally, inefficient system-level mechanisms, such as autoscaling and scheduling, can cause further delays~\cite{pusztai2022polaris}. Environmental factors such as device heterogeneity and network congestion can cause increased execution queue and/or failures~\cite{scf}. Therefore, a performance model that captures broader influencing factors is essential for identifying performance bottlenecks and creating effective solutions to address them.

\paragraph{RC-2: How do serverless runtimes impact function performance in the Edge-Cloud Continuum?}

Wasm provides advantages such as compact binary size, which contributes to reduced cold start latency compared to container-based runtimes. However, Wasm executes within a sandboxed environment with linear memory and no direct access to host resources, relying on the WASI to enable interactions with the host for filesystem and network access, introducing overhead, particularly in data-intensive applications~\cite{ContainerRuntimeYet2025,Cwasi2023}. Furthermore, while Wasm interpreter-based execution ensures portability across platforms, it incurs higher runtime overhead compared to AoT and Just-in-Time (JIT) compilation~\cite{CrossArchitectureWasmECC,WasmNextGen}. In contrast, container-based runtimes benefit from direct host integration, relying on Linux namespaces and cgroups to enforce isolation. While container isolation provides portability and scalability, it potentially exposes both the host and the functions to threats posed by untrusted tenants~\cite{PushingWasmEdge,ChallengesServerless2019}. Therefore, it is crucial to understand the factors that influence serverless function performance under varying workloads, systems, and environmental characteristics.

\paragraph{RC-3: How to characterize serverless function performance under different workload characteristics and heterogeneous environments, such as the Edge-Cloud Continuum?}

Existing performance characterization either focuses on standalone Wasm runtimes or serverless functions without capturing all influencing performance characteristics of the Edge-Cloud Continuum~\cite{wang_how_far_2022,WasmNextGen,CrossArchitectureWasmECC,EvaluatingWasmEdge}. Evaluating serverless runtimes requires the ability to simulate concurrency, diverse workload characteristics (e.g., compute- vs. data-intensive), cold vs. warm execution, and BaaS access. Additionally, fine-grained telemetry must be collected to isolate compute, I/O, and serialization time, and to attribute resource usage to specific functions. Therefore, a modular and extensible benchmarking that integrates multiple runtimes, enabling reproducible experiments with precise resource attribution, is essential for a precise performance characterization.

\section{Serverless Performance Model for the Edge-Cloud Continuum}\label{sec:methodology}
\begin{figure}[t]
    \centering
    \includegraphics[width=0.8\linewidth]{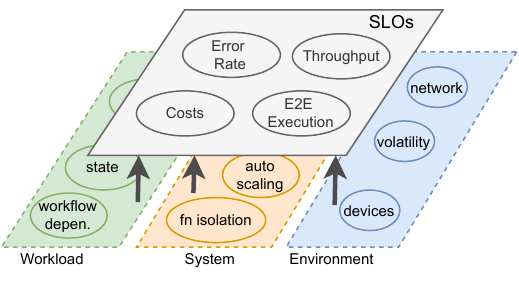}
    \caption{Serverless Performance Model for the Edge-Cloud Continuum, highlighting performance factors: workload, system, and environment. }
    \label{fig:performance_model}
\end{figure}

Typically, the performance of serverless functions is measured by SLOs, such as end-to-end execution time, throughput, costs, and error rate. However, the performance of serverless functions can be influenced by various factors. To effectively analyze this variability, we introduce a performance model that categorizes the main performance factors into: workload, system, and environment, as shown in \cref{fig:performance_model}. Each dimension represents a unique influence layer in the function lifecycle (\cref{fig:function_lifecycle}) that affects the overall function performance. Lumos' performance model facilitates the profiling of serverless functions' performance, allowing for precise benchmarking.

\subsection{Workload Performance Factors}

\paragraph{Function Heterogeneity} Serverless functions display a variety of workload characteristics, such as compute-intensive and data-intensive tasks that involve significant data movement. Lumos offers benchmarking capabilities, enabling users to evaluate how the performance of runtimes scales with different application variations, providing users with insights, as certain runtimes may perform better with specific workloads, such as CPU-bound tasks, while struggling with others, such as I/O operations that involve a high overhead from system calls.

\paragraph{Workflow Dependencies} Serverless workflows often comprise interconnected functions that exchange ephemeral data\cite{goldfish2024}. Performance issues in one function in the workflow may impact the performance of the entire workflow, as subsequent functions depend on outputs from previous functions.

\paragraph{Function State and BaaS}  Typically, serverless functions are stateless and rely on external remote services to store function state.  These additional services used by the functions are referred to as Backend-as-a-Service (BaaS). The integration of BaaS goes beyond just state management; serverless functions also utilize BaaS to perform more complex tasks, such as ETL and AI inference. While BaaS accelerates development and promotes modularity, it introduces external dependencies and increases latency and network overhead~\cite{sand,faastlane,Truffle2024}.

\paragraph{Serialization Overhead} Serverless functions are typically written in high-level languages such as NodeJS, Python, and Java and exchange complex data structures, which incur high (de)serialization costs~\cite{state-of-serverless,datadog2023state,serialization-free}. For instance, AWS@Lambda has reported over 67\% of functions running in less than 20ms. On the other hand, the (de)serialization process of a complex Python object of ~3MB takes around 10 ms. Therefore, as serverless functions are short-lived, additional overhead with (de)serialization can significantly impact the execution time~\cite{serialization-free,faastlane}. 

\subsection{System Performance Factors}

\paragraph{Performance Isolation} In a multi-tenant environment, such as serverless, performance isolation ensures that one function’s behavior does not impact others. Poor isolation performance results in noisy neighbor effects, where co-located functions contend for CPU, memory, or network bandwidth. For instance, a compute-bound function may monopolize CPU time, leading to scheduling delays for co-resident functions. Similarly, shared memory bandwidth and last-level cache can be saturated by data-intensive functions, causing delays on neighbor functions~\cite{xavier2015performance,zhou2017performance,xie2021evaluation}. 

\paragraph{Function Isolation} Serverless functions are tiny pieces of code wrapped in bundles and operated by serverless platforms. Typically, serverless functions are executed in a multi-tenant environment, sharing host hardware and OS. To ensure that functions do not interfere with each other, they are isolated using different mechanisms, such as Docker containers, unikernels, and Wasm VMs~\cite{faasm,sledge,PushingWasmEdge}. Container-based isolation offers native execution of functions but increases cold starts as functions must provision their file system and setup cgroups. WasmVM offers a lightweight isolation, but it introduces overhead due to WASI communication~\cite{Cwasi2023,ContainerRuntimeYet2025,WasmExecutionModel}. 

\paragraph{Function Invocation} Typically, serverless function execution is characterized by dynamic invocation behaviors such as bursty invocations, scheduled triggers, and multi-tenant requests~\cite{state-of-serverless}. 
Bursty workloads can temporarily overload the platform, leading to request queuing, increased cold starts, and resource contention. Additionally, uneven invocation rates can create a concurrency imbalance, where dominant functions consume shared resources, ultimately increasing the latency of identical functions up to 338\%~\cite{wen2025unveiling}. To maintain performance under diverse workloads, platforms must balance their system-level mechanisms, such as concurrency limits and scaling thresholds.

\paragraph{Auto Scaling and Cold Starts} In serverless computing, functions are not continuously running and are instantiated only upon demand. When no active instance is available, the platform must provision the necessary runtime environment, a process known as a cold start. To manage demand, auto scalers dynamically adjust the number of function instances based on incoming traffic~\cite{scf,serverless-why-when-how,2024selfprovisioningInfrastructure}. While this elasticity improves resource utilization, it is sensitive to feedback lag. Bursty workloads may not be reflected in time, potentially causing the auto scaler to react too late or scale inadequately~\cite{wang2018peeking}.

\begin{figure}[t]
    \centering
    \includegraphics[width=\linewidth]{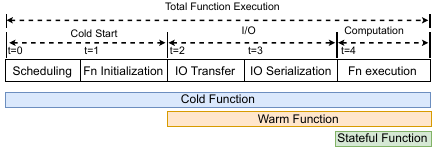}
    \caption{Serverless function lifecycle highlighting cold, warm, and stateful functions}
    \label{fig:function_lifecycle}
\end{figure}

\subsection{Environmental Performance Factors}

\paragraph{Nodes and Devices Capabilities} The performance of serverless functions is influenced by the characteristics of their execution environments. Specifically, in the Edge-Cloud Continuum, which can vary due to device heterogeneity, including differences in CPU architecture, memory capacity, and GPU capabilities, functions that require specific hardware may experience degraded performance if deployed on under-provisioned nodes. Similarly, memory-bound functions can suffer from allocation failures on low-memory edge devices, leading to increased latency or function retries.

\paragraph{Network Variability}
Typically, serverless functions rely on network connectivity to access data, invoke downstream services, or communicate with external BaaS components. In the Edge–Cloud Continuum, function performance is affected by variable latency, fluctuating bandwidth, and intermittent connectivity \cite{Databelt2025}.. High jitter or packet loss can lead to retries, timeout errors, and degraded throughput Therefore, network performance directly impacts total execution time, particularly for data-intensive functions.

\paragraph{Devices Volatility} In resource-constrained environments such as Edge-Cloud Continuum, devices might be battery-powered or change positions, making nodes' availability unpredictable and consequently leading to load shifts to other nodes in the cluster. This instability leads the platform to reschedule functions, which may incur cold starts, increased communication overhead, and consequently impact overall performance.

\section{Lumos Architecture Overview and Performance Metrics}\label{sec:architecture}
In this section, we present Lumos architecture overview and performance metrics that enable serverless performance characterization.

\begin{figure}[t]
    \centering
    \includegraphics[width=\linewidth]{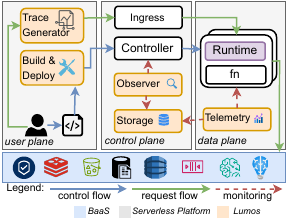}
    \caption{Lumos Architecture Overview}
    \label{fig:architecture}
\end{figure}

\subsection{Lumos Architecture Overview}

Lumos provides an end-to-end benchmarking tool across three planes, as shown in (\cref{fig:architecture}). In the \emph{user plane}, functions are prepared and reproducible traces are generated; in the \emph{control plane}, scheduling and routing are observed and control metadata is persisted; and in the \emph{data plane}, state I/O is instrumented to capture compute and BaaS effects. This three-plane decomposition provides detailed performance insights while ensuring both portability and low overhead. It maintains a clear separation of concerns, allowing workloads to evolve independently from the serverless platform. Additionally, it enables seamless runtime changes (such as Wasm or containers) without necessitating changes to the measurement stack. 
As each step in a function’s execution presents unique trade-offs (e.g, cold starts and BaaS access latency), Lumos profiles these function execution steps to identify bottlenecks and measure their impacts across serverless workloads.

\paragraph{Built-in Functions} Lumos includes a set of representative serverless workloads designed to capture diverse performance behaviors. These functions cover both data-intensive and compute-bound workloads. They are used to evaluate how different serverless runtimes handle varying workload characteristics. \cref{tab:evaluation_workloads_sorted} provides an overview of the selected workloads and their corresponding profiles.

\subsubsection{Lumos Components} Lumos is composed of six components that span from control to data plane and user plane.

\paragraph{Build and Deploy} Lumos provides a modular build pipeline that supports reproducible and portable benchmarking. The build pipeline enables serverless functions to be compiled into Docker containers using a base image or Wasm modules in interpreted or AoT compilation. 
The deployment pipeline integrates seamlessly with the serverless platform, such as Knative. Lumos provides deployment templates that enable consistent rollout across runtime configurations while isolating experiment conditions such as storage type, payload size, and function parameters. Based on the user input, Lumos selects the runtime, selection via the orchestrator deployment config, i.e., \texttt{.spec.template.spec.runtimeClassName}. 

\paragraph{Trace Generator} It creates synthetic invocation traces based on configurable profiles. The generator can simulate concurrency levels, delayed invocations, and realistic bursty invocations.

\paragraph{Observer} It intercepts and timestamps key lifecycle events (e.g., ingress, data IO, serialization, function ready, response sent). It aggregates metrics from runtime logs, platform APIs, and custom probes injected into the function code to aid the data IO and serialization identification. Additionally, the observer collects the function runtime ID and its PID to enable precise correlation across logs and telemetry streams. 

\paragraph{Storage} It stores collected metrics, logs, and structured traces. The storage enables future integration with observability tools such as Prometheus, Loki, or OpenTelemetry for unified analysis and monitoring across experiments.

\paragraph{Telemetry} It collects fine-grained resource metrics (e.g., CPU times, memory usage, I/O operations) via cgroup, and function runtime, thus allowing the identification of resource usage for specific functions. It tracks usage to specific processes using the container IDs, enabling per-request resource attribution.

\begin{table}[t]
\centering
\caption{Description of Lumos Built-in Functions}
\resizebox{\linewidth}{!}{%
\begin{tabular}{llcc}
\toprule
Function            & Description                 & High CPU & High I/O   \\ \midrule
\rowcolor[HTML]{EFEFEF} 
audio-generation    & Generate audio              & \cmark   & \cmark   \\
fuzzy-search        & Fuzzy text search           & \cmark   & \cmark   \\
\rowcolor[HTML]{EFEFEF}
language-detection  & Text language detect        & \cmark   & \cmark   \\
encrypt-message     & Encrypt data                & \cmark   & \xmark   \\
\rowcolor[HTML]{EFEFEF}
decrypt-message     & Decrypt data                & \cmark   & \xmark   \\
fibonacci           & Fibonacci sequence          & \cmark   & \xmark   \\
\rowcolor[HTML]{EFEFEF}
prime-numbers       & Prime number calculation    & \cmark   & \xmark   \\
mandelbrot-bitmap   & Generate Mandelbrot bitmap  & \cmark   & \cmark   \\
\bottomrule
\end{tabular}%
}
\label{tab:evaluation_workloads_sorted}
\end{table}

\subsection{Performance Metrics}
In this section, we define the metrics provided by Lumos that enable granular performance evaluation of serverless functions. Each metric captures a distinct aspect of the function lifecycle, shown in ~\cref{fig:function_lifecycle}, from invocation to execution and resource consumption, enabling detailed analysis across different serverless runtimes.

\paragraph{Total Function Execution Time} It reflects the combined impact of platform delays such as auto scaling, scheduling, request forwarding, as well as function logic, I/O operations, and other external dependencies. It is the most visible metric to users and essential for evaluating non-functional requirements such as SLO compliance.

\paragraph{I/O Latency} It shows the time to access external services such as databases and object stores. In serverless workflows, frequent access to BaaS can dominate execution time, especially in data-intensive applications or workflows with function chains.

\paragraph{Serialization} It accounts for the cost of encoding and decoding data between functions or when communicating with external services. Complex data structures incur significant (de)serialization overhead, which impacts short-lived functions significantly.

\paragraph{Compute Time} It measures the function computation without cold start initialization, IO, and serialization overhead. The compute time isolates the pure logic execution phase to enable runtime efficiency evaluation, highlighting performance differences in compute-bound workloads across serverless runtimes.

\paragraph{Image Size} Function image size impacts the pull time, which is necessary during cold start. Specifically, in the congested network, larger images may suffer increased pull times, impacting overall function execution time. On the other hand, smaller images, especially in Wasm runtimes, can significantly reduce cold starts.

\paragraph{Coldstart Time} It measures the time between a request's arrival on the platform and the function becoming ready for execution. It includes image loading, scheduling, and environment provisioning. 

\paragraph{CPU Usage} It provides fine-grained CPU time consumed during function execution. It is essential for profiling compute-bound functions and assessing runtime efficiency across different execution environments.

\paragraph{Memory Usage}  It measures the memory footprint of the function. It is particularly important for data-intensive functions in constrained environments such as edge nodes, where memory limits may affect function placement.

\section{Implementation}\label{sec:implementation}

Lumos is implemented as a modular benchmarking framework designed to evaluate the performance of serverless runtimes such as Wasm and container-based serverless functions across heterogeneous edge platforms. Lumos components are implemented in Bash, Python, and Rust, and support full automation of build, deployment, benchmarking, and telemetry collection pipelines. It is published as an open-source framework and is available on GitHub\footnote{\url{https://github.com/polaris-slo-cloud/lumos}}

\paragraph{Configurability and Logging}
All configuration settings for the benchmarking pipeline are centralized in a global file, including properties such as architecture targets, image registry endpoints, and run. Function-specific parameters such as input sizes and storage types are encoded via JSON payloads. During runtime, each function communicates with its backend-as-a-service storage based on runtime-configurable parameters passed via JSON payloads. This design allows precise analysis of the impact of data and compute-intensive workloads on cold and warm starts.

\paragraph{Functions and Workloads}
The Lumos built-in workloads include eight serverless functions implemented in Rust, selected to represent data-intensive and compute-intensive functions. Each function exposes a single POST endpoint and supports native and Wasm compilation targets via conditional Cargo configurations. Functions are invoked using structured payloads, with parameters controlling function behavior and input size. \cref{tab:evaluation_workloads_sorted} summarizes each workload and resource profile.

\section{Evaluation}\label{sec:evaluation}

The goal of our experiments is to characterize performance under varying input sizes and concurrency levels and assess resource efficiency. We benchmark the serverless functions spanning data and compute-intensive workloads as described in \cref{tab:evaluation_workloads_sorted}. Input sizes range from 512KB to 2MB for data-intensive and up to 2M integer as described in \cref{tab:overview_cpu_payload_sizes}, reflecting structured data commonly exchanged between serverless functions in edge–cloud workflows~\cite{state-of-serverless}.
We conduct both sequential executions, as well as parallel executions,  aligning with real-world serverless workflows, as outlined in~\cite{CloudProgrammingSimplified}.

\begin{table}[t!]
\centering
\caption{Payload size groups}
\resizebox{0.75\linewidth}{!}{%
\begin{tabular}{llll}
\toprule
\shortstack{Function} & \shortstack{Group Size 1} & \shortstack{Group Size 2} & \shortstack{Group Size 3} \\ 
\midrule
\rowcolor[HTML]{EFEFEF}
audio-generation & 512 Kb & 1024 Kb & 2048 Kb \\
fuzzy-search & 512 Kb & 1024 Kb & 2048 Kb \\
\rowcolor[HTML]{EFEFEF}
language-detection & 512 Kb & 1024 Kb & 2048 Kb \\
encrypt-message & 512 Kb & 1024 Kb & 2048 Kb \\
\rowcolor[HTML]{EFEFEF}
decrypt-message & 512 Kb & 1024 Kb & 2048 Kb \\
fibonacci & 10,000 int & 100,000 int & 1,000,000 int \\
\rowcolor[HTML]{EFEFEF}
prime-numbers & 10,000 int & 100,000 int & 1,000,000 int \\
mandelbrot-bitmap & 512 Kb & 1024 Kb & 2048 Kb \\
\bottomrule
\end{tabular}%
}
\label{tab:overview_cpu_payload_sizes}
\end{table}

\subsection{Experimental Setup}

Our experiments are conducted on a heterogeneous edge cluster to emulate realistic resource-constrained environments. Our testbed includes two Raspberry Pi Model 5B with 4x2.4 GHz CPUs with 16 GB and 8 GB RAM, respectively, and two Raspberry Pi 4B with 4x1.8 GHz CPUs and 4 GB RAM. All devices have 64 GB storage and 100 Mbps network links. 
As an orchestration tool, we use MicroK8s, Knative as a serverless platform, and Redis as storage. For the Wasm runtime management, we leverage the Kwasm operator. Each serverless function is deployed as a Knative service.  All serverless functions are implemented in Rust. Telemetry for CPU and memory usage is collected using \texttt{telemd}~\cite{telemd}, which runs on each node. Each experiment is repeated 10 times, and mean values are reported to ensure statistical significance.

\begin{figure}[t]
    \centering
    \begin{tikzpicture}
        \begin{axis}[
            ybar,
            bar width=6pt,
            width=\linewidth,
            height=3.5cm,
            ymax=125,
            ylabel={Image Size (MB)},
            ylabel style={font=\scriptsize},
            xticklabel style={font=\tiny, align=center},
            symbolic x coords={Audio\\Generation,Fuzzy\\Search,Language\\Detection,Encrypt\\Message,Decrypt\\Message,Fibonacci, Prime\\Numbers, Mandelbrot\\Bitmap},
            enlarge x limits=0.08,
            xtick=data,
            grid=none,
            tick align=inside,
            legend style={at={(0.5,1.25)}, anchor=north, legend columns=-1, font=\scriptsize, draw=none},
            legend image code/.code={
              \draw[#1, fill] (0cm,-0.1cm) rectangle (0.3cm,0.15cm);
            },
            nodes near coords,
            every node near coord/.append style={
                font=\tiny,
                yshift=8pt,
                xshift=5pt,
                rotate=90
            }
        ]
    
        \addplot[fill=blue!60, postaction={pattern=crosshatch}] coordinates {
            (Audio\\Generation, 81.8)
            (Fuzzy\\Search, 92.8)
            (Language\\Detection, 82)
            (Encrypt\\Message, 81.7)
            (Decrypt\\Message, 81.7)
            (Fibonacci, 80.7)
            (Prime\\Numbers, 80.7)
            (Mandelbrot\\Bitmap, 81.7)
        };
        \addlegendentry{Container}
    
        \addplot[fill=green!60!black, postaction={pattern=north west lines}] coordinates {
            (Audio\\Generation, 1.8)
            (Fuzzy\\Search, 1.89)
            (Language\\Detection, 2.08)
            (Encrypt\\Message, 1.82)
            (Decrypt\\Message, 1.82)
            (Fibonacci, 1.27)
            (Prime\\Numbers, 1.27)
            (Mandelbrot\\Bitmap, 1.84)
        };
        \addlegendentry{Wasm}

        \addplot[fill=orange!80, postaction={pattern=crosshatch dots}] coordinates {
            (Audio\\Generation, 3.51)
            (Fuzzy\\Search, 3.68)
            (Language\\Detection, 3.88)
            (Encrypt\\Message, 3.54)
            (Decrypt\\Message, 3.54)
            (Fibonacci, 2.55)
            (Prime\\Numbers, 2.55)
            (Mandelbrot\\Bitmap, 3.58)
        };
        \addlegendentry{Wasm AoT}
    
        \end{axis}
    \end{tikzpicture}

    \caption{Image sizes of serverless functions}
    \label{fig:image_size}
\end{figure}
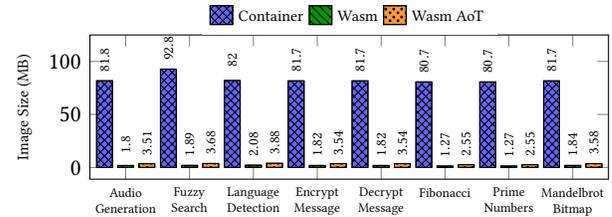

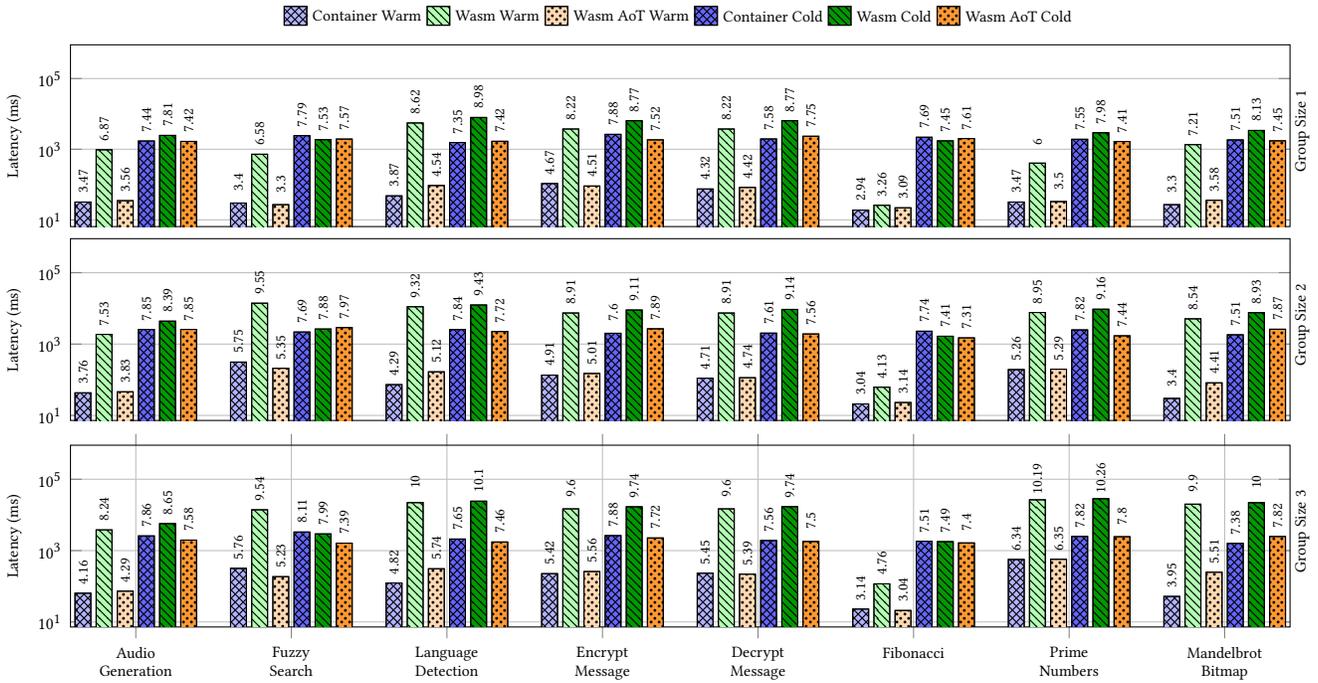
\begin{figure*}[t]
    \centering
    \begin{tikzpicture}
        \begin{axis}[
            ybar,
            bar width=6pt,
            width=\textwidth,
            height=4cm,
            ymax=900000,
            ymode=log,
            ylabel={Latency (ms)},
            xlabel={ Group Size 1},
            xlabel style={
                at={(1.01,0.8)}, 
                rotate=90, 
                font=\scriptsize 
            },
            ylabel style={font=\scriptsize},
            yticklabel style={font=\scriptsize},
            xticklabel style={font=\scriptsize, align=center},
            symbolic x coords={Audio\\Generation,Fuzzy\\Search,Language\\Detection,Encrypt\\Message,Decrypt\\Message,Fibonacci, Prime\\Numbers, Mandelbrot\\Bitmap},
            xtick=\empty,
            enlarge x limits=0.06,
            grid=major,
            legend style={at={(0.5,1.25)}, anchor=north, legend columns=-1, font=\scriptsize, draw=none},
            legend image code/.code={
              \draw[#1, fill] (0cm,-0.1cm) rectangle (0.3cm,0.15cm);
            },
            nodes near coords,
            every node near coord/.append style={
                font=\tiny,
                yshift=8pt,
                xshift=5pt,
                rotate=90
            }
        ]
    
        \addplot[fill=blue!30, postaction={pattern=crosshatch}] coordinates {
            (Audio\\Generation, 32)
            (Fuzzy\\Search, 30)
            (Language\\Detection, 48)
            (Encrypt\\Message, 107)
            (Decrypt\\Message, 75)
            (Fibonacci, 19)
            (Prime\\Numbers, 32)
            (Mandelbrot\\Bitmap, 27)
        };
        \addlegendentry{Container Warm}

        \addplot[fill=green!30, postaction={pattern=north west lines}] coordinates {
            (Audio\\Generation, 963)
            (Fuzzy\\Search, 721)
            (Language\\Detection, 5537)
            (Encrypt\\Message, 3733)
            (Decrypt\\Message, 3730)
            (Fibonacci, 26)
            (Prime\\Numbers, 404)
            (Mandelbrot\\Bitmap, 1351)
        };
        \addlegendentry{Wasm Warm}

        \addplot[fill=orange!30, postaction={pattern=crosshatch dots}] coordinates {
            (Audio\\Generation, 35)
            (Fuzzy\\Search, 27)
            (Language\\Detection, 94)
            (Encrypt\\Message, 91)
            (Decrypt\\Message, 83)
            (Fibonacci, 22)
            (Prime\\Numbers, 33)
            (Mandelbrot\\Bitmap, 36)
        };
        \addlegendentry{Wasm AoT Warm}
    
        \addplot[fill=blue!60, postaction={pattern=crosshatch}] coordinates {
            (Audio\\Generation, 1708)
            (Fuzzy\\Search, 2428)
            (Language\\Detection, 1552)
            (Encrypt\\Message, 2634)
            (Decrypt\\Message, 1951)
            (Fibonacci, 2183)
            (Prime\\Numbers, 1910)
            (Mandelbrot\\Bitmap, 1835)
        };
        \addlegendentry{Container Cold}
    
        \addplot[fill=green!60!black, postaction={pattern=north west lines}] coordinates {
            (Audio\\Generation, 2462)
            (Fuzzy\\Search, 1865)
            (Language\\Detection, 7914)
            (Encrypt\\Message, 6407)
            (Decrypt\\Message, 6414)
            (Fibonacci, 1724)
            (Prime\\Numbers, 2923)
            (Mandelbrot\\Bitmap, 3399)
        };
        \addlegendentry{Wasm Cold}

        \addplot[fill=orange!80, postaction={pattern=crosshatch dots}] coordinates {
            (Audio\\Generation, 1664)
            (Fuzzy\\Search, 1937)
            (Language\\Detection, 1676)
            (Encrypt\\Message, 1842)
            (Decrypt\\Message, 2328)
            (Fibonacci, 2009)
            (Prime\\Numbers, 1657)
            (Mandelbrot\\Bitmap, 1727)
        };
        \addlegendentry{Wasm AoT Cold}
    
        \end{axis}
    \end{tikzpicture}

    \begin{tikzpicture}
        \begin{axis}[
            ybar,
            bar width=6pt,
            width=\textwidth,
            height=4cm,
            ymode=log,
            ymax=900000,
            ylabel={Latency (ms)},
            xlabel={ Group Size 2},
            xlabel style={
                at={(1.01,0.8)}, 
                rotate=90, 
                font=\scriptsize 
            },
            ylabel style={font=\scriptsize},
            yticklabel style={font=\scriptsize},
            xticklabel style={font=\scriptsize, align=center},
            symbolic x coords={Audio\\Generation,Fuzzy\\Search,Language\\Detection,Encrypt\\Message,Decrypt\\Message,Fibonacci, Prime\\Numbers, Mandelbrot\\Bitmap},
            xtick=\empty,
            enlarge x limits=0.06,
            grid=major,
            legend style={at={(0.5,1.25)}, anchor=north, legend columns=-1, font=\scriptsize, draw=none},
            legend image code/.code={
                \draw[##1, fill=##1] (0cm,-0.1cm) rectangle (0.3cm,0.15cm);
            },
            nodes near coords,
            every node near coord/.append style={
                font=\tiny,
                yshift=8pt,
                xshift=5pt,
                rotate=90
            }
        ]
            
        \addplot[fill=blue!30, postaction={pattern=crosshatch}] coordinates {
            (Audio\\Generation, 43)
            (Fuzzy\\Search, 313)
            (Language\\Detection, 73)
            (Encrypt\\Message, 135)
            (Decrypt\\Message, 111)
            (Fibonacci, 21)
            (Prime\\Numbers, 193)
            (Mandelbrot\\Bitmap, 30)
        };
    
        \addplot[fill=green!30, postaction={pattern=north west lines}] coordinates {
            (Audio\\Generation, 1871)
            (Fuzzy\\Search, 13989)
            (Language\\Detection, 11134)
            (Encrypt\\Message, 7433)
            (Decrypt\\Message, 7420)
            (Fibonacci, 62)
            (Prime\\Numbers, 7686)
            (Mandelbrot\\Bitmap, 5100)
        };

        \addplot[fill=orange!30, postaction={pattern=crosshatch dots}] coordinates {
            (Audio\\Generation, 46)
            (Fuzzy\\Search, 210)
            (Language\\Detection, 167)
            (Encrypt\\Message, 150)
            (Decrypt\\Message, 114)
            (Fibonacci, 23)
            (Prime\\Numbers, 198)
            (Mandelbrot\\Bitmap, 82)
        };
         
        \addplot[fill=blue!60, postaction={pattern=crosshatch}] coordinates {
            (Audio\\Generation, 2576)
            (Fuzzy\\Search, 2179)
            (Language\\Detection, 2548)
            (Encrypt\\Message, 2008)
            (Decrypt\\Message, 2028)
            (Fibonacci, 2292)
            (Prime\\Numbers, 2498)
            (Mandelbrot\\Bitmap, 1822)
        };
    
        \addplot[fill=green!60!black, postaction={pattern=north west lines}] coordinates {
            (Audio\\Generation, 4392)
            (Fuzzy\\Search, 2644)
            (Language\\Detection, 12465)
            (Encrypt\\Message, 9018)
            (Decrypt\\Message, 9357)
            (Fibonacci, 1657)
            (Prime\\Numbers, 9482)
            (Mandelbrot\\Bitmap, 7578)
        };
       \addplot[fill=orange!80, postaction={pattern=crosshatch dots}] coordinates {
            (Audio\\Generation, 2578)
            (Fuzzy\\Search, 2906)
            (Language\\Detection, 2254)
            (Encrypt\\Message, 2681)
            (Decrypt\\Message, 1921)
            (Fibonacci, 1493)
            (Prime\\Numbers, 1706)
            (Mandelbrot\\Bitmap, 2605)
        };
    
        \end{axis}
    \end{tikzpicture}

    \begin{tikzpicture}
        \begin{axis}[
            ybar,
            bar width=6pt,
            width=\textwidth,
            height=4cm,
            ymode=log,
            ymax=900000,
            ylabel={Latency (ms)},
            xlabel={ Group Size 3},
            xlabel style={
                at={(1.01,0.8)}, 
                rotate=90, 
                font=\scriptsize 
            },
            ylabel style={font=\scriptsize},
            yticklabel style={font=\scriptsize},
            xticklabel style={font=\scriptsize, align=center},
            symbolic x coords={Audio\\Generation,Fuzzy\\Search,Language\\Detection,Encrypt\\Message,Decrypt\\Message,Fibonacci, Prime\\Numbers, Mandelbrot\\Bitmap},
            xtick=data,
            enlarge x limits=0.06,
            grid=major,
            legend style={at={(0.5,1.25)}, anchor=north, legend columns=-1, font=\scriptsize, draw=none},
            legend image code/.code={
                \draw[##1, fill=##1] (0cm,-0.1cm) rectangle (0.3cm,0.15cm);
            },
            nodes near coords,
            every node near coord/.append style={
                font=\tiny,
                yshift=8pt,
                xshift=5pt,
                rotate=90
            }
        ]

        \addplot[fill=blue!30, postaction={pattern=crosshatch}] coordinates {
            (Audio\\Generation, 64)
            (Fuzzy\\Search, 318)
            (Language\\Detection, 124)
            (Encrypt\\Message, 226)
            (Decrypt\\Message, 233)
            (Fibonacci, 23)
            (Prime\\Numbers, 565)
            (Mandelbrot\\Bitmap, 52)
        };

        \addplot[fill=green!30, postaction={pattern=north west lines}] coordinates {
            (Audio\\Generation, 3806)
            (Fuzzy\\Search, 13927)
            (Language\\Detection, 22095)
            (Encrypt\\Message, 14779)
            (Decrypt\\Message, 14780)
            (Fibonacci, 117)
            (Prime\\Numbers, 26559)
            (Mandelbrot\\Bitmap, 20003)
        };

        \addplot[fill=orange!30, postaction={pattern=crosshatch dots}] coordinates {
            (Audio\\Generation, 73)
            (Fuzzy\\Search, 186)
            (Language\\Detection, 310)
            (Encrypt\\Message, 259)
            (Decrypt\\Message, 219)
            (Fibonacci, 21)
            (Prime\\Numbers, 572)
            (Mandelbrot\\Bitmap, 248)
        };
    
        \addplot[fill=blue!60, postaction={pattern=crosshatch}] coordinates {
            (Audio\\Generation, 2584)
            (Fuzzy\\Search, 3319)
            (Language\\Detection, 2108)
            (Encrypt\\Message, 2634)
            (Decrypt\\Message, 1919)
            (Fibonacci, 1832)
            (Prime\\Numbers, 2502)
            (Mandelbrot\\Bitmap, 1600)
        };

        \addplot[fill=green!60!black, postaction={pattern=north west lines}] coordinates {
            (Audio\\Generation, 5719)
            (Fuzzy\\Search, 2940)
            (Language\\Detection, 24287)
            (Encrypt\\Message, 16917)
            (Decrypt\\Message, 17022)
            (Fibonacci, 1785)
            (Prime\\Numbers, 28474)
            (Mandelbrot\\Bitmap, 22031)
        };
        \addplot[fill=orange!80, postaction={pattern=crosshatch dots}] coordinates {
            (Audio\\Generation, 1968)
            (Fuzzy\\Search, 1619)
            (Language\\Detection, 1743)
            (Encrypt\\Message, 2263)
            (Decrypt\\Message, 1801)
            (Fibonacci, 1642)
            (Prime\\Numbers, 2450)
            (Mandelbrot\\Bitmap, 2483)
        };

        \end{axis}
    \end{tikzpicture}

    \caption{Execution time comparison across I/O- and CPU-intensive functions for varying input sizes and group sizes.}
    \label{fig:combined_exec_io_cpu}
\end{figure*}

\begin{figure}[t]
    \centering
    
        \begin{tikzpicture}
            \begin{axis}[
                hide axis,
                xmin=0, xmax=1,
                ymin=0, ymax=1,
                legend style={
                    draw=none,
                    font=\footnotesize,
                    legend columns=-1,
                    column sep=1ex
                }
            ]
                \addlegendimage{mark=o, blue, thick,mark size=1pt}
                \addlegendentry{Runc}

                \addlegendimage{mark=x, thick, green!60!black,mark size=1pt}
                \addlegendentry{Wasm}

                \addlegendimage{mark=triangle, thick, orange,mark size=1pt}
                \addlegendentry{Wasm AoT}
    
            \end{axis}
        \end{tikzpicture}
            
    \resizebox{\linewidth}{!}{
        \begin{tikzpicture}
            \begin{axis}[
                width=0.19\textwidth,
                height=3cm,
                ylabel={\shortstack{Data IO\\CDF}},
                xlabel style={align=center,font=\footnotesize},
                ylabel style={font=\footnotesize},
                grid=both
            ]
                \addplot[mark=o,thick, blue, mark size=1pt] coordinates {
                (0.793, 0.2000) (0.793, 0.4000) (3.459, 0.6000) (4.360, 0.8000) (5.535, 1.0000)
                };
                \addplot[ mark=x, thick, green!60!black,mark size=1pt] coordinates {
                 (2.390, 0.2000) (2.735, 0.4000) (4.957, 0.6000) (6.868, 0.8000) (7.064, 1.0000)
                };

                 \addplot[mark=x, thick, orange,mark size=1pt] coordinates {
                (0.431, 0.2000) (0.560, 0.4000) (0.693, 0.6000) (3.742, 0.8000) (5.369, 1.0000)
                };
            \end{axis}
        \end{tikzpicture}
        \begin{tikzpicture}
            \begin{axis}[
                width=0.19\textwidth,
                height=3cm,
                ytick=\empty,
                xlabel style={align=center,font=\footnotesize},
                ylabel style={font=\footnotesize},
                grid=both
            ]
                \addplot[mark=o,thick, blue, mark size=1pt] coordinates {
                (0.983, 0.2000) (2.553, 0.4000) (4.468, 0.6000) (4.513, 0.8000) (8.301, 1.0000)
                };
                
                \addplot[mark=x, thick, green!60!black,mark size=1pt] coordinates {
                (2.609, 0.2000) (2.710, 0.4000) (5.866, 0.6000) (7.506, 0.8000) (12.711, 1.0000)
                };

                 \addplot[mark=x, thick, orange,mark size=1pt] coordinates {
                (0.482, 0.2000) (0.970, 0.4000) (2.829, 0.6000) (4.929, 0.8000) (9.854, 1.0000)
                };
            \end{axis}
        \end{tikzpicture}
        \begin{tikzpicture}
            \begin{axis}[
                width=0.19\textwidth,
                height=3cm,
                ytick=\empty,
                ymode=log,
                xlabel style={align=center,font=\footnotesize},
                ylabel style={font=\footnotesize},
                grid=both
            ]
                \addplot[mark=o,thick, blue, mark size=1pt] coordinates {
                (0.809, 0.2000) (2.626, 0.4000) (4.492, 0.6000) (5.907, 0.8000) (8.141, 1.0000)
                };
                \addplot[mark=x, thick, green!60!black,mark size=1pt] coordinates {
               (2.477, 0.2000) (2.836, 0.4000) (5.792, 0.6000) (7.703, 0.8000) (10.007, 1.0000)
                };

                 \addplot[mark=x, thick, orange,mark size=1pt] coordinates {
                (0.452, 0.2000) (0.702, 0.4000) (2.631, 0.6000) (4.266, 0.8000) (7.446, 1.0000)
                };
            \end{axis}
        \end{tikzpicture}
        \begin{tikzpicture}
            \begin{axis}[
                width=0.19\textwidth,
                height=3cm,
                ytick=\empty,
                ymode=log,
                xlabel style={align=center,font=\footnotesize},
                ylabel style={font=\footnotesize},
                grid=both
            ]
                \addplot[mark=o,thick, blue, mark size=1pt] coordinates {
                (0.997, 0.2000) (3.629, 0.4000) (5.759, 0.6000) (11.752, 0.8000) (13.392, 1.0000)
                };
                \addplot[mark=x, thick, green!60!black,mark size=1pt] coordinates {
                (2.477, 0.2000) (2.898, 0.4000) (6.451, 0.6000) (8.042, 0.8000) (13.157, 1.0000)
                };

                 \addplot[mark=x, thick, orange,mark size=1pt] coordinates {
                (0.423, 0.2000) (0.770, 0.4000) (3.654, 0.6000) (5.828, 0.8000) (10.487, 1.0000)
                };
            \end{axis}
        \end{tikzpicture}
    }

    \resizebox{\linewidth}{!}{

        \begin{tikzpicture}
            \begin{axis}[
                width=0.19\textwidth,
                height=3cm,
                ylabel={\shortstack{Serialization\\CDF}},
                xlabel style={align=center,font=\footnotesize},
                ylabel style={font=\footnotesize},
                xlabel={Fuzzy\\Search},
                grid=both
            ]
                \addplot[mark=o,thick, blue, mark size=1pt] coordinates {
                (0.003, 0.2000) (0.007, 0.4000) (0.007, 0.6000) (0.187, 0.8000) (0.197, 1.0000)
                };
                \addplot[mark=x, thick, green!60!black,mark size=1pt] coordinates {
                (0.202, 0.2000) (1.418, 0.4000) (1.536, 0.6000) (27.130, 0.8000) (27.943, 1.0000)
                };

                \addplot[mark=x, thick, orange,mark size=1pt] coordinates {
                (0.005, 0.2000) (0.013, 0.4000) (0.014, 0.6000) (0.239, 0.8000) (0.276, 1.0000)
                };
            \end{axis}
        \end{tikzpicture}

        \begin{tikzpicture}
            \begin{axis}[
                width=0.19\textwidth,
                height=3cm,
                ytick=\empty,
                ymode=log,
                xlabel style={align=center,font=\footnotesize},
                ylabel style={font=\footnotesize},
                xlabel={Language\\Detection},
                grid=both
            ]
                \addplot[mark=o,thick, blue, mark size=1pt] coordinates {
                (0.004, 0.2000) (0.043, 0.4000) (0.446, 0.6000) (0.937, 0.8000) (1.607, 1.0000)
                };
                
                \addplot[mark=x, thick, green!60!black,mark size=1pt] coordinates {
                (0.316, 0.2000) (4.247, 0.4000) (40.474, 0.6000) (78.458, 0.8000) (155.401, 1.0000)
                };

                \addplot[mark=x, thick, orange,mark size=1pt] coordinates {
                (0.006, 0.2000) (0.046, 0.4000) (0.442, 0.6000) (0.901, 0.8000) (1.789, 1.0000)
                };
            \end{axis}
        \end{tikzpicture}

        \begin{tikzpicture}
            \begin{axis}[
                width=0.19\textwidth,
                height=3cm,
                ytick=\empty,
                ymode=log,
                xlabel style={align=center,font=\footnotesize},
                ylabel style={font=\footnotesize},
                xlabel={Encrypt\\Message},
                grid=both
            ]
                \addplot[mark=o,thick, blue, mark size=1pt] coordinates {
                (0.004, 0.2000) (0.044, 0.4000) (0.445, 0.6000) (0.932, 0.8000) (1.925, 1.0000)
                };
                \addplot[mark=x, thick, green!60!black,mark size=1pt] coordinates {
               (0.265, 0.2000) (3.952, 0.4000) (40.742, 0.6000) (77.589, 0.8000) (159.620, 1.0000)
                };

                \addplot[mark=x, thick, orange,mark size=1pt] coordinates {
                (0.007, 0.2000) (0.050, 0.4000) (0.427, 0.6000) (1.552, 0.8000) (1.731, 1.0000)
                };
            \end{axis}
        \end{tikzpicture}


        \begin{tikzpicture}
            \begin{axis}[
                width=0.19\textwidth,
                height=3cm,
                ytick=\empty,
                ymode=log,
                xlabel style={align=center,font=\footnotesize},
                ylabel style={font=\footnotesize},
                xlabel={Decrypt\\Message},
                grid=both
            ]
                \addplot[mark=o,thick, blue, mark size=1pt] coordinates {
                 (0.003, 0.2000) (0.008, 0.4000) (0.130, 0.6000) (0.247, 0.8000) (0.403, 1.0000)
                };
                \addplot[mark=x, thick, green!60!black,mark size=1pt] coordinates {
                (0.207, 0.2000) (2.037, 0.4000) (18.869, 0.6000) (36.231, 0.8000) (73.114, 1.0000)
                };

                \addplot[mark=x, thick, orange,mark size=1pt] coordinates {
                 (0.005, 0.2000) (0.029, 0.4000) (0.158, 0.6000) (0.327, 0.8000) (0.693, 1.0000)
                };
            \end{axis}
        \end{tikzpicture}

    }

    \caption{CDF Latency IO and Serialization Overhead}
    \label{fig:serialization}
\end{figure}
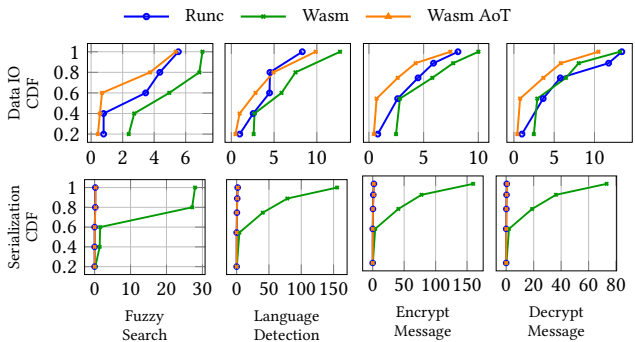

\subsection{Baselines}

For our baselines, we use RunC as the container runtime for native functions, and WasmEdge as the WebAssembly runtime, evaluated in both interpreted and AoT modes. While we evaluate functions on the WasmEdge runtime, Lumos is designed to be extensible and modular, allowing seamless integration of alternative WebAssembly runtimes such as Wasmtime and Spin via Kwasm.

\subsection{Image Size Results}

In this experiment, we compare the image sizes of deployed serverless functions across the three runtime configurations: Docker Container, WasmEdge (interpreted), and WasmEdge AoT. Image size is a key factor in cold start performance, deployment latency, and storage overhead, particularly relevant in edge environments where bandwidth and storage are constrained. Each workload is packaged using identical source code and dependencies. While containers have a minimal base image (i.e., \texttt{debian:bullseye-slim}), WebAssembly images have a single binary packed from scratch.

\cref{fig:image_size} presents the resulting image sizes. Docker Container images produce significantly larger images, averaging over 80MB, AoT-compiled binaries ranging from 2.5 to 3.9 MB, and interpreted Wasm binaries range from 1.2 to 2MB. Wasm shows a reduction in image size up to 30x compared to containers, highlighting a major advantage of Wasm for scenarios with limited network capacity.

\begin{tcolorbox}[colframe=gray, colback=gray!10, coltitle=black, boxrule=0.3mm,  rounded corners,left=1mm, right=1mm, top=1mm, bottom=1mm]
\textbf{\emph{Finding 1:}}  Wasm reduces image size by up to 30x compared to containers, highlighting its benefit in scenarios with limited network.
\end{tcolorbox}

\subsection{Function Execution Time Results}\label{subsec:latency}

\cref{fig:combined_exec_io_cpu} compares the end-to-end function execution time (in milliseconds) of multiple serverless functions under warm and cold starts across three group sizes, reflecting increasing input payloads as defined in \cref{tab:overview_cpu_payload_sizes}. 
Across all group sizes, Container and WasmEdge AoT show lower latency under warm starts, while interpreted WasmEdge exhibits significantly higher latency, particularly in compute-intensive functions such as Prime Numbers and Mandelbrot Bitmap. For cold starts, WebAssembly AoT performs competitively with containers, especially for smaller group sizes, though performance diverges as the input size increases. Interpreted WasmEdge suffers from consistently high cold start latency, with values exceeding 24 seconds in large-input compute-intensive workloads.

\input{evaluation/res_usage}
\input{evaluation/scalability}

\begin{tcolorbox}[colframe=gray, colback=gray!10, coltitle=black, boxrule=0.3mm,  rounded corners,left=1mm, right=1mm, top=1mm, bottom=1mm]
\textbf{\emph{Finding 2:}} \textbf{Warm Functions:} Wasm AoT shows up to 1.1x the latency of containers across both compute- and data-intensive functions.
Wasm interpreted is up to 30x slower for compute-intensive functions and up to 55x slower for data-intensive functions.
\end{tcolorbox}

\begin{tcolorbox}[colframe=gray, colback=gray!10, coltitle=black, boxrule=0.3mm,  rounded corners,left=1mm, right=1mm, top=1mm, bottom=1mm]
\textbf{\emph{Finding 3:}}
\textbf{Cold Functions:} Wasm AoT shows up to 16\% lower latency than containers, particularly in data-intensive functions.
Wasm interpreted remains up to 1.7x slower for compute-intensive and up to 2.9x slower for data-intensive functions.
\end{tcolorbox}

\subsection{IO and Serialization Results}
As Wasm showed increased execution time, particularly for data-intensive functions (\cref{subsec:latency}), in this experiment, we isolate the impact of I/O and serialization overhead to understand how data handling contributes to the overall performance of different runtimes.
\cref{fig:serialization} shows the normalized data retrieval and serialization overhead across increasing payload sizes. Wasm AoT introduces up to 2x I/O overhead compared to containers, while Wasm interpreted incurs up to 7x higher I/O overhead, especially for large payloads. During serialization, Wasm AoT maintains close-to-native serialization performance, with only up to 2x overhead over containers, while Wasm interpreted shows serialization costs exceeding 10x for larger payloads.

\begin{tcolorbox}[colframe=gray, colback=gray!10, coltitle=black, boxrule=0.3mm,  rounded corners,left=1mm, right=1mm, top=1mm, bottom=1mm]
    \textbf{\emph{Finding 4:}} I/O and serialization significantly impact performance in Wasm runtimes. Wasm AoT incurs up to 2x in I/O and serialization overhead, while Wasm interpreted shows up to 7x higher I/O overhead and over 10x serialization overhead compared to containers.
\end{tcolorbox}

\subsection{Resource Usage Results}

\cref{fig:res_usage} compares normalized CPU and memory usage under varying input sizes, across different serverless runtimes, Container, WasmEdge (interpreted), and WasmEdge AoT runtimes. Across all workloads, containers consistently exhibit the lowest CPU and memory consumption. For compute-intensive functions such as Fibonacci and Prime Numbers, WasmEdge incurs 5–20x higher CPU usage than containers at larger input sizes, while AoT reduces the overhead but remains up to 3x higher. Memory usage patterns reveal similar trends, where interpreted WasmEdge consumes up to 42 MB for Fuzzy Search and Language Detection, compared to under 1 MB for containers. AoT shows a lower memory usage ranging from 5 to 16 MB.

\begin{tcolorbox}[colframe=gray, colback=gray!10, coltitle=black, boxrule=0.3mm,  rounded corners,left=1mm, right=1mm, top=1mm, bottom=1mm]
    \textbf{\emph{Finding 5:}} Wasm shows up to 20x higher CPU usage and 40x more memory, for both compute- and data-intensive workloads. 

\end{tcolorbox}

\subsection{Scalability Results}

\cref{fig:scalability} presents normalized request per second (RPS), i.e.,  throughput over latency for increasing concurrency levels across eight serverless workloads ranging from 1 to 200 RPS. The results compare three runtime configurations: Container, WasmEdge (interpreted), and WasmEdge AoT. Containers maintain the most stable performance under load, sustaining throughput even as latency exceeds 10s. Data-intensive functions such as Audio Generation and Fuzzy Search show up to 50\% lower latency in Container compared to WasmEdge at peak load. Compute-intensive workloads such as Prime Numbers and Mandelbrot Bitmap reveal even greater differences: WasmEdge experiences high degradation up to 461s latency, while workloads running on containers complete execution below 18s. AoT compilation improves WasmEdge's performance under moderate load, but high concurrency still leads to substantial latency growth, particularly in mixed workloads such as Language Detection, where interpreted WasmEdge peaks at over 930s.

\begin{tcolorbox}[colframe=gray, colback=gray!10, coltitle=black, boxrule=0.3mm,  rounded corners,left=1mm, right=1mm, top=1mm, bottom=1mm]
    \textbf{\emph{Finding 6:}} Workloads running Containers consistently outperform WebAssembly (interpreted), achieving up to 27x lower latency and 2x higher throughput under high concurrency. WasmEdge AoT improves over interpreted mode but still degrades under high load.
    \textbf{Compute-intensive workloads}  show the largest gaps, with WebAssembly with quick performance degradation. 
    \textbf{Data-intensive functions} show a fair benefit from WebAssembly AoT compilation, which narrows the performance gap to containers but still lags under high concurrency.
\end{tcolorbox}

\section{Related Work}\label{sec:related_work}

\paragraph{Serverless Function Benchmarks}
SeBS~\cite{SBS_copik2021,SeBS20,sebs-flow} introduces a benchmarking suite for evaluating the performance of serverless workflows across cloud platforms, measuring function execution time, orchestration overhead, cold start frequency, scalability, and cost. SeBS provides a platform-independent methodology to compare serverless workflow performance. However, SeBS does not support multiple serverless runtimes, which is crucial for resource-constrained environments such as the Edge-Cloud Continuum. 
XFBench~\cite{XFBench} evaluates serverless platforms using customizable workflows, workloads, and functions. It generates and deploys function DAGs, invoking workloads via JMeter, and collecting telemetry for detailed analysis across multiple platforms. XFBench enables reproducible benchmarking and helps developers analyze FaaS suitability, optimize workflow composition, and identify cloud platform bottlenecks. EdgeFaaSBench~\cite{EdgeFaaSBench} evaluates the performance of serverless computing on edge devices with heterogeneous hardware capabilities, such as CPU and GPU. EdgeFaaSBench captures metrics such as cold/warm start times, concurrent execution overheads, and system-level utilization. However, EdgeFaaSBench does not differentiate the impact of workload characteristics, such as compute and data-intensive. Thus, it limits the user's ability to optimize the workflow based on the workload characteristics.

\paragraph{WebAssembly Benchmarks}
WABench~\cite{wang_how_far_2022} presents a comprehensive characterization of predominant standalone WebAssembly runtimes, including Wasmtime, Wasmer, WAVM, Wasm3, and WAMR.  While WABench focuses on low-level runtime behavior (e.g., performance, instruction count, cache usage, memory overhead) using synthetic benchmarks and full applications, it lacks consideration of serverless-specific factors such as cold vs. warm starts. Lumos captures the trade-offs of using WebAssembly in serverless computing in heterogeneous environments such as the Edge-Cloud Continuum rather than standalone Wasm runtimes in isolation. In ~\cite{PerformanceWasmCloser25}, authors investigate the performance implications of integrating Wasm into container runtimes across diverse hardware platforms, emphasizing image size, startup time, and container pull time, but do not consider workload characteristics, such as workload data IO, serialization, or WASI overhead. 
In ~\cite{BenchmarkingWasm2025}, the authors present a comprehensive benchmark of Wasm for embedded systems, analyzing performance, memory use, and runtime features. They show that WASM, especially with AoT compilation, offers competitive performance and memory efficiency.
Liu et. al.~\cite{ContainerRuntimeYet2025} evaluate the performance of Wasm-based container runtimes in comparison to traditional Docker containers and standalone Wasm runtimes. Liu shows that Wasm containers currently fail to deliver performance advantages over Docker containers due to overheads introduced by WASI. However, Liu's benchmark only considers Wasm in interpreter mode, which suffers from performance overhead. In Lumos, we identify the WASI overhead in interpreter and in AoT compilation, providing insightful data on how to improve performance in cases where WASI overhead is a bottleneck.
In ~\cite{WasmExecutionModel}, the authors propose an execution model for serverless functions at the edge by replacing containers with Wasm modules. Although authors show that wasm-based serverless functions suffer performance overhead compared to native containers, they do not identify application characteristics that directly impact performance. 

In contrast to the presented state-of-the-art approaches, Lumos evaluates Wasm and container runtimes across diverse workload types (CPU-and I/O-bound) under varying invocation patterns and in heterogeneous environments, also capturing the performance overhead from workloads that leverage WASI, enabling a more comprehensive performance characterization of WebAssembly as a serverless runtime in the Edge-Cloud Continuum.

\section{Conclusion \& Future Work}\label{sec:conclusion}
In this paper, we introduced Lumos, a benchmarking tool and performance model for evaluating serverless runtimes in the Edge–Cloud Continuum. Lumos provides a performance characterization of serverless workloads with different characteristics, such as compute- and data-intensive, by analyzing fine-grained metrics such as total function execution time, cold and warm starts, as well as IO and serialization latency. Our experiments provide a benchmark of Wasm and container as serverless runtimes, evaluating their performance based on workload, environment, and system factors.
Our results show that Wasm still incurs notable latency overhead, especially in interpreted mode, where latency can increase by up to 27x compared to containers. Additionally, we further observe that Wasm incurs up to 7× higher I/O and up to 10× higher serialization overhead, posing significant challenges for data-intensive workloads, which can limit Wasm's benefits in such scenarios.

In the future, we plan to extend Lumos to enable the performance characterization of serverless functions with real-world traces and in different environments, such as the 3D Computing Continuum that unifies Edge-Cloud and Space. Lumos will allow users to identify performance bottlenecks and evaluate function trade-offs in edge and orbital environments without requiring real satellite deployments.

\section*{Acknowledgment}
This work is partially funded by the Austrian Research Promotion Agency (FFG) under the project RapidREC (Project No. 903884).
This work has received funding from the Austrian Internet Stiftung under the NetIdee project LEO Trek (ID~7442).
Partly funded by EU Horizon Europe under GA no. 101070186 (TEADAL) and GA no. 101192912 (NexaSphere).


\bibliographystyle{ACM-Reference-Format}
\bibliography{references}


\begin{thebibliography}{44}


\ifx \showCODEN    \undefined \def \showCODEN     #1{\unskip}     \fi
\ifx \showDOI      \undefined \def \showDOI       #1{#1}\fi
\ifx \showISBNx    \undefined \def \showISBNx     #1{\unskip}     \fi
\ifx \showISBNxiii \undefined \def \showISBNxiii  #1{\unskip}     \fi
\ifx \showISSN     \undefined \def \showISSN      #1{\unskip}     \fi
\ifx \showLCCN     \undefined \def \showLCCN      #1{\unskip}     \fi
\ifx \shownote     \undefined \def \shownote      #1{#1}          \fi
\ifx \showarticletitle \undefined \def \showarticletitle #1{#1}   \fi
\ifx \showURL      \undefined \def \showURL       {\relax}        \fi
\providecommand\bibfield[2]{#2}
\providecommand\bibinfo[2]{#2}
\providecommand\natexlab[1]{#1}
\providecommand\showeprint[2][]{arXiv:#2}

\bibitem[Akkus et~al\mbox{.}(2018)]%
        {sand}
\bibfield{author}{\bibinfo{person}{Istemi~Ekin Akkus}, \bibinfo{person}{Ruichuan Chen}, \bibinfo{person}{Ivica Rimac}, {et~al\mbox{.}}} \bibinfo{year}{2018}\natexlab{}.
\newblock \showarticletitle{{SAND}: Towards {High-Performance} Serverless Computing}. In \bibinfo{booktitle}{\emph{2018 USENIX Annual Technical Conference (USENIX ATC 18)}}. \bibinfo{publisher}{USENIX Association}, \bibinfo{address}{Boston, MA}, \bibinfo{pages}{923--935}.
\newblock


\bibitem[Copik et~al\mbox{.}(2025)]%
        {SeBS20}
\bibfield{author}{\bibinfo{person}{Marcin Copik}, \bibinfo{person}{Alexandru Calotoiu}, {and} \bibinfo{person}{Torsten Hoefler}.} \bibinfo{year}{2025}\natexlab{}.
\newblock \showarticletitle{SeBS 2.0: Keeping up with the Clouds}. In \bibinfo{booktitle}{\emph{Proceedings of the 3rd Workshop on SErverless Systems, Applications and MEthodologies}} (Rotterdam, Netherlands) \emph{(\bibinfo{series}{SESAME' 25})}. \bibinfo{pages}{42–44}.
\newblock
\showISBNx{9798400715570}
\urldef\tempurl%
\url{https://doi.org/10.1145/3721465.3721867}
\showDOI{\tempurl}


\bibitem[Copik et~al\mbox{.}(2021)]%
        {SBS_copik2021}
\bibfield{author}{\bibinfo{person}{Marcin Copik}, \bibinfo{person}{Grzegorz Kwasniewski}, \bibinfo{person}{Maciej Besta}, {et~al\mbox{.}}} \bibinfo{year}{2021}\natexlab{}.
\newblock \showarticletitle{SeBS: A Serverless Benchmark Suite for Function-as-a-Service Computing}. In \bibinfo{booktitle}{\emph{Proceedings of the 22nd International Middleware Conference}} (Qu\'{e}bec city, Canada) \emph{(\bibinfo{series}{Middleware '21})}. \bibinfo{pages}{64–78}.
\newblock
\showISBNx{9781450385343}
\urldef\tempurl%
\url{https://doi.org/10.1145/3464298.3476133}
\showDOI{\tempurl}


\bibitem[{Datadog}(2023)]%
        {datadog2023state}
\bibfield{author}{\bibinfo{person}{{Datadog}}.} \bibinfo{year}{2023}\natexlab{}.
\newblock \bibinfo{title}{The State of Serverless}.
\newblock \bibinfo{howpublished}{\url{https://www.datadoghq.com/state-of-serverless/}}.
\newblock


\bibitem[{Edgerun}(2025)]%
        {telemd}
\bibfield{author}{\bibinfo{person}{{Edgerun}}.} \bibinfo{year}{2025}\natexlab{}.
\newblock \bibinfo{title}{{Telemd}}.
\newblock
\newblock
\urldef\tempurl%
\url{https://github.com/edgerun/telemd}
\showURL{%
\tempurl}


\bibitem[Eismann et~al\mbox{.}(2022)]%
        {state-of-serverless}
\bibfield{author}{\bibinfo{person}{Simon Eismann}, \bibinfo{person}{Joel Scheuner}, \bibinfo{person}{Erwin~van Eyk}, {et~al\mbox{.}}} \bibinfo{year}{2022}\natexlab{}.
\newblock \showarticletitle{The State of Serverless Applications: Collection, Characterization, and Community Consensus}.
\newblock \bibinfo{journal}{\emph{IEEE Transactions on Software Engineering}} \bibinfo{volume}{48}, \bibinfo{number}{10} (\bibinfo{year}{2022}), \bibinfo{pages}{4152--4166}.
\newblock
\urldef\tempurl%
\url{https://doi.org/10.1109/TSE.2021.3113940}
\showDOI{\tempurl}


\bibitem[Eismann et~al\mbox{.}(2021)]%
        {serverless-why-when-how}
\bibfield{author}{\bibinfo{person}{Simon Eismann}, \bibinfo{person}{Joel Scheuner}, \bibinfo{person}{Erwin van Eyk}, {et~al\mbox{.}}} \bibinfo{year}{2021}\natexlab{}.
\newblock \showarticletitle{Serverless Applications: Why, When, and How?}
\newblock \bibinfo{journal}{\emph{IEEE Software}} \bibinfo{volume}{38}, \bibinfo{number}{1} (\bibinfo{year}{2021}), \bibinfo{pages}{32--39}.
\newblock
\urldef\tempurl%
\url{https://doi.org/10.1109/MS.2020.3023302}
\showDOI{\tempurl}


\bibitem[Gackstatter et~al\mbox{.}(2022)]%
        {PushingWasmEdge}
\bibfield{author}{\bibinfo{person}{Philipp Gackstatter}, \bibinfo{person}{Pantelis~A. Frangoudis}, {and} \bibinfo{person}{Schahram Dustdar}.} \bibinfo{year}{2022}\natexlab{}.
\newblock \showarticletitle{Pushing Serverless to the Edge with WebAssembly Runtimes}. In \bibinfo{booktitle}{\emph{2022 22nd IEEE International Symposium on Cluster, Cloud and Internet Computing (CCGrid)}}. \bibinfo{pages}{140--149}.
\newblock
\urldef\tempurl%
\url{https://doi.org/10.1109/CCGrid54584.2022.00023}
\showDOI{\tempurl}


\bibitem[Gadepalli et~al\mbox{.}(2020)]%
        {sledge}
\bibfield{author}{\bibinfo{person}{Phani~Kishore Gadepalli}, \bibinfo{person}{Sean McBride}, \bibinfo{person}{Gregor Peach}, {et~al\mbox{.}}} \bibinfo{year}{2020}\natexlab{}.
\newblock \showarticletitle{Sledge: a Serverless-first, Light-weight Wasm Runtime for the Edge}. In \bibinfo{booktitle}{\emph{Proceedings of the 21st International Middleware Conference}} (Delft, Netherlands) \emph{(\bibinfo{series}{Middleware '20})}. \bibinfo{pages}{265–279}.
\newblock
\showISBNx{9781450381536}
\urldef\tempurl%
\url{https://doi.org/10.1145/3423211.3425680}
\showDOI{\tempurl}


\bibitem[Gadepalli et~al\mbox{.}(2019)]%
        {ChallengesServerless2019}
\bibfield{author}{\bibinfo{person}{Phani~Kishore Gadepalli}, \bibinfo{person}{Gregor Peach}, \bibinfo{person}{Ludmila Cherkasova}, \bibinfo{person}{Rob Aitken}, {and} \bibinfo{person}{Gabriel Parmer}.} \bibinfo{year}{2019}\natexlab{}.
\newblock \showarticletitle{Challenges and Opportunities for Efficient Serverless Computing at the Edge}. In \bibinfo{booktitle}{\emph{2019 38th Symposium on Reliable Distributed Systems (SRDS)}}. \bibinfo{pages}{261--2615}.
\newblock
\urldef\tempurl%
\url{https://doi.org/10.1109/SRDS47363.2019.00036}
\showDOI{\tempurl}


\bibitem[Hall and Ramachandran(2019)]%
        {WasmExecutionModel}
\bibfield{author}{\bibinfo{person}{Adam Hall} {and} \bibinfo{person}{Umakishore Ramachandran}.} \bibinfo{year}{2019}\natexlab{}.
\newblock \showarticletitle{An execution model for serverless functions at the edge}. In \bibinfo{booktitle}{\emph{Proceedings of the International Conference on Internet of Things Design and Implementation}} (Montreal, Quebec, Canada) \emph{(\bibinfo{series}{IoTDI '19})}. \bibinfo{pages}{225–236}.
\newblock
\showISBNx{9781450362832}
\urldef\tempurl%
\url{https://doi.org/10.1145/3302505.3310084}
\showDOI{\tempurl}


\bibitem[Jangda et~al\mbox{.}(2019)]%
        {jangda_not-so-fast_2019}
\bibfield{author}{\bibinfo{person}{Abhinav Jangda}, \bibinfo{person}{Bobby Powers}, \bibinfo{person}{Emery~D. Berger}, {and} \bibinfo{person}{Arjun Guha}.} \bibinfo{year}{2019}\natexlab{}.
\newblock \showarticletitle{Not So Fast: Analyzing the Performance of {WebAssembly} vs. Native Code}. In \bibinfo{booktitle}{\emph{2019 USENIX Annual Technical Conference (USENIX ATC 19)}}. \bibinfo{publisher}{USENIX Association}, \bibinfo{address}{Renton, WA}, \bibinfo{pages}{107--120}.
\newblock
\showISBNx{978-1-939133-03-8}


\bibitem[Jiang et~al\mbox{.}(2024)]%
        {RevealingWasmPerformance}
\bibfield{author}{\bibinfo{person}{Shuyao Jiang}, \bibinfo{person}{Ruiying Zeng}, \bibinfo{person}{Zihao Rao}, {et~al\mbox{.}}} \bibinfo{year}{2024}\natexlab{}.
\newblock \showarticletitle{Revealing Performance Issues in Server-Side WebAssembly Runtimes via Differential Testing}. In \bibinfo{booktitle}{\emph{Proceedings of the 38th IEEE/ACM International Conference on Automated Software Engineering}} (Echternach, Luxembourg) \emph{(\bibinfo{series}{ASE '23})}. \bibinfo{publisher}{IEEE Press}, \bibinfo{pages}{661–672}.
\newblock
\showISBNx{9798350329964}
\urldef\tempurl%
\url{https://doi.org/10.1109/ASE56229.2023.00088}
\showDOI{\tempurl}


\bibitem[Jonas et~al\mbox{.}(2019)]%
        {CloudProgrammingSimplified}
\bibfield{author}{\bibinfo{person}{Eric Jonas}, \bibinfo{person}{Johann Schleier-Smith}, {et~al\mbox{.}}} \bibinfo{year}{2019}\natexlab{}.
\newblock \bibinfo{title}{Cloud Programming Simplified: A Berkeley View on Serverless Computing}.
\newblock
\newblock
\showeprint[arxiv]{1902.03383}~[cs.OS]


\bibitem[Kakati and Brorsson(2024)]%
        {CrossArchitectureWasmECC}
\bibfield{author}{\bibinfo{person}{Sangeeta Kakati} {and} \bibinfo{person}{Mats Brorsson}.} \bibinfo{year}{2024}\natexlab{}.
\newblock \showarticletitle{A Cross-Architecture Evaluation of WebAssembly in the Cloud-Edge Continuum}. In \bibinfo{booktitle}{\emph{2024 IEEE 24th International Symposium on Cluster, Cloud and Internet Computing (CCGrid)}}. \bibinfo{pages}{337--346}.
\newblock
\urldef\tempurl%
\url{https://doi.org/10.1109/CCGrid59990.2024.00046}
\showDOI{\tempurl}


\bibitem[Kakati and Brorsson(2025)]%
        {PerformanceWasmCloser25}
\bibfield{author}{\bibinfo{person}{Sangeeta Kakati} {and} \bibinfo{person}{Mats Brorsson}.} \bibinfo{year}{2025}\natexlab{}.
\newblock \showarticletitle{Performance and Usability Implications of Multiplatform and WebAssembly Containers}. In \bibinfo{booktitle}{\emph{Proceedings of the 15th International Conference on Cloud Computing and Services Science - Volume 1: CLOSER}}. INSTICC, \bibinfo{publisher}{SciTePress}, \bibinfo{pages}{15--25}.
\newblock
\showISBNx{978-989-758-747-4}
\urldef\tempurl%
\url{https://doi.org/10.5220/0013203200003950}
\showDOI{\tempurl}


\bibitem[Kjorveziroski and Filiposka(2023a)]%
        {WasmNextGen}
\bibfield{author}{\bibinfo{person}{Vojdan Kjorveziroski} {and} \bibinfo{person}{Sonja Filiposka}.} \bibinfo{year}{2023}\natexlab{a}.
\newblock \showarticletitle{WebAssembly as an Enabler for Next Generation Serverless Computing}.
\newblock \bibinfo{journal}{\emph{Journal of Grid Computing}} \bibinfo{volume}{21}, \bibinfo{number}{3} (\bibinfo{year}{2023}), \bibinfo{pages}{34}.
\newblock
\showISSN{1572-9184}
\urldef\tempurl%
\url{https://doi.org/10.1007/s10723-023-09669-8}
\showDOI{\tempurl}


\bibitem[Kjorveziroski and Filiposka(2023b)]%
        {WebAssemblyOrchestration}
\bibfield{author}{\bibinfo{person}{Vojdan Kjorveziroski} {and} \bibinfo{person}{Sonja Filiposka}.} \bibinfo{year}{2023}\natexlab{b}.
\newblock \showarticletitle{WebAssembly Orchestration in the Context of Serverless Computing}.
\newblock \bibinfo{journal}{\emph{Journal of Network and Systems Management}} \bibinfo{volume}{31}, \bibinfo{number}{3} (\bibinfo{year}{2023}), \bibinfo{pages}{62}.
\newblock
\showISSN{1573-7705}
\urldef\tempurl%
\url{https://doi.org/10.1007/s10922-023-09753-0}
\showDOI{\tempurl}


\bibitem[Kotni et~al\mbox{.}(2021)]%
        {faastlane}
\bibfield{author}{\bibinfo{person}{Swaroop Kotni}, \bibinfo{person}{Ajay Nayak}, \bibinfo{person}{Vinod Ganapathy}, {and} \bibinfo{person}{Arkaprava Basu}.} \bibinfo{year}{2021}\natexlab{}.
\newblock \showarticletitle{Faastlane: Accelerating {Function-as-a-Service} Workflows}. In \bibinfo{booktitle}{\emph{2021 USENIX Annual Technical Conference (USENIX ATC 21)}}. \bibinfo{publisher}{USENIX Association}, \bibinfo{pages}{805--820}.
\newblock
\showISBNx{978-1-939133-23-6}


\bibitem[Kulkarni et~al\mbox{.}(2024)]%
        {XFBench}
\bibfield{author}{\bibinfo{person}{Varad Kulkarni}, \bibinfo{person}{Nikhil Reddy}, \bibinfo{person}{Tuhin Khare}, {et~al\mbox{.}}} \bibinfo{year}{2024}\natexlab{}.
\newblock \showarticletitle{XFBench: A Cross-Cloud Benchmark Suite for Evaluating FaaS Workflow Platforms}. In \bibinfo{booktitle}{\emph{2024 IEEE 24th International Symposium on Cluster, Cloud and Internet Computing (CCGrid)}}. \bibinfo{pages}{543--556}.
\newblock
\urldef\tempurl%
\url{https://doi.org/10.1109/CCGrid59990.2024.00067}
\showDOI{\tempurl}


\bibitem[Liu et~al\mbox{.}(2025)]%
        {ContainerRuntimeYet2025}
\bibfield{author}{\bibinfo{person}{Mugeng Liu}, \bibinfo{person}{Haiyang Shen}, \bibinfo{person}{Yixuan Zhang}, {et~al\mbox{.}}} \bibinfo{year}{2025}\natexlab{}.
\newblock \showarticletitle{WebAssembly for Container Runtime: Are We There Yet?}
\newblock \bibinfo{journal}{\emph{ACM Trans. Softw. Eng. Methodol.}} (\bibinfo{date}{Feb.} \bibinfo{year}{2025}).
\newblock
\showISSN{1049-331X}
\urldef\tempurl%
\url{https://doi.org/10.1145/3712197}
\showDOI{\tempurl}


\bibitem[Liu et~al\mbox{.}(2023)]%
        {TheGapResearchRealWorld}
\bibfield{author}{\bibinfo{person}{Qingyuan Liu}, \bibinfo{person}{Dong Du}, \bibinfo{person}{Yubin Xia}, \bibinfo{person}{Ping Zhang}, {and} \bibinfo{person}{Haibo Chen}.} \bibinfo{year}{2023}\natexlab{}.
\newblock \showarticletitle{The Gap Between Serverless Research and Real-World Systems}. In \bibinfo{booktitle}{\emph{Proceedings of the 2023 ACM Symposium on Cloud Computing}} (Santa Cruz, CA, USA) \emph{(\bibinfo{series}{SoCC '23})}. \bibinfo{pages}{475–485}.
\newblock
\showISBNx{9798400703874}
\urldef\tempurl%
\url{https://doi.org/10.1145/3620678.3624785}
\showDOI{\tempurl}


\bibitem[Lu et~al\mbox{.}(2024)]%
        {serialization-free}
\bibfield{author}{\bibinfo{person}{Fangming Lu}, \bibinfo{person}{Xingda Wei}, \bibinfo{person}{Zhuobin Huang}, \bibinfo{person}{Rong Chen}, \bibinfo{person}{Minyu Wu}, {and} \bibinfo{person}{Haibo Chen}.} \bibinfo{year}{2024}\natexlab{}.
\newblock \showarticletitle{Serialization/Deserialization-free State Transfer in Serverless Workflows}. In \bibinfo{booktitle}{\emph{Proceedings of the Nineteenth European Conference on Computer Systems}} (Athens, Greece) \emph{(\bibinfo{series}{EuroSys '24})}. \bibinfo{publisher}{Association for Computing Machinery}, \bibinfo{address}{New York, NY, USA}, \bibinfo{pages}{132–147}.
\newblock
\showISBNx{9798400704376}
\urldef\tempurl%
\url{https://doi.org/10.1145/3627703.3629568}
\showDOI{\tempurl}


\bibitem[Marcelino et~al\mbox{.}(2025a)]%
        {Databelt2025}
\bibfield{author}{\bibinfo{person}{Cynthia Marcelino}, \bibinfo{person}{Leonard Guelmino}, \bibinfo{person}{Thomas Pusztai}, {and} \bibinfo{person}{Stefan Nastic}.} \bibinfo{year}{2025}\natexlab{a}.
\newblock \showarticletitle{Databelt: A continuous data path for serverless workflows in the 3D compute continuum}.
\newblock \bibinfo{journal}{\emph{Journal of Systems Architecture}} (\bibinfo{year}{2025}).
\newblock
\showISSN{1383-7621}
\urldef\tempurl%
\url{https://doi.org/10.1016/j.sysarc.2025.103577}
\showDOI{\tempurl}


\bibitem[Marcelino and Nastic(2024a)]%
        {Cwasi2023}
\bibfield{author}{\bibinfo{person}{Cynthia Marcelino} {and} \bibinfo{person}{Stefan Nastic}.} \bibinfo{year}{2024}\natexlab{a}.
\newblock \showarticletitle{CWASI: A WebAssembly Runtime Shim for Inter-function Communication in the Serverless Edge-Cloud Continuum}. In \bibinfo{booktitle}{\emph{Proceedings of the Eighth ACM/IEEE Symposium on Edge Computing}} (Wilmington, DE, USA) \emph{(\bibinfo{series}{SEC '23})}. \bibinfo{pages}{158–170}.
\newblock
\showISBNx{9798400701238}
\urldef\tempurl%
\url{https://doi.org/10.1145/3583740.3626611}
\showDOI{\tempurl}


\bibitem[Marcelino and Nastic(2024b)]%
        {Truffle2024}
\bibfield{author}{\bibinfo{person}{Cynthia Marcelino} {and} \bibinfo{person}{Stefan Nastic}.} \bibinfo{year}{2024}\natexlab{b}.
\newblock \showarticletitle{Truffle: Efficient Data Passing for Data-Intensive Serverless Workflows in the Edge-Cloud Continuum}. In \bibinfo{booktitle}{\emph{2024 IEEE/ACM 17th International Conference on Utility and Cloud Computing (UCC)}}. \bibinfo{pages}{53--62}.
\newblock
\urldef\tempurl%
\url{https://doi.org/10.1109/UCC63386.2024.00017}
\showDOI{\tempurl}


\bibitem[Marcelino et~al\mbox{.}(2025b)]%
        {goldfish2024}
\bibfield{author}{\bibinfo{person}{Cynthia Marcelino}, \bibinfo{person}{Jack Shahhoud}, {and} \bibinfo{person}{Stefan Nastic}.} \bibinfo{year}{2025}\natexlab{b}.
\newblock \showarticletitle{GoldFish: Serverless Actors with Short-Term Memory State for the Edge-Cloud Continuum}. In \bibinfo{booktitle}{\emph{Proceedings of the 14th International Conference on the Internet of Things}} \emph{(\bibinfo{series}{IoT '24})}. \bibinfo{pages}{56–64}.
\newblock
\showISBNx{9798400712852}
\urldef\tempurl%
\url{https://doi.org/10.1145/3703790.3703797}
\showDOI{\tempurl}


\bibitem[Mendki(2020)]%
        {EvaluatingWasmEdge}
\bibfield{author}{\bibinfo{person}{Pankaj Mendki}.} \bibinfo{year}{2020}\natexlab{}.
\newblock \showarticletitle{Evaluating Webassembly Enabled Serverless Approach for Edge Computing}. In \bibinfo{booktitle}{\emph{2020 IEEE Cloud Summit}}. \bibinfo{pages}{161--166}.
\newblock
\urldef\tempurl%
\url{https://doi.org/10.1109/IEEECloudSummit48914.2020.00031}
\showDOI{\tempurl}


\bibitem[Moron and Wallentowitz(2025)]%
        {BenchmarkingWasm2025}
\bibfield{author}{\bibinfo{person}{Konrad Moron} {and} \bibinfo{person}{Stefan Wallentowitz}.} \bibinfo{year}{2025}\natexlab{}.
\newblock \showarticletitle{Benchmarking WebAssembly for Embedded Systems}.
\newblock \bibinfo{journal}{\emph{ACM Trans. Archit. Code Optim.}} (\bibinfo{date}{May} \bibinfo{year}{2025}).
\newblock
\showISSN{1544-3566}
\urldef\tempurl%
\url{https://doi.org/10.1145/3736169}
\showDOI{\tempurl}


\bibitem[Ménétrey et~al\mbox{.}(2022)]%
        {wasmCommonLayer}
\bibfield{author}{\bibinfo{person}{Jämes Ménétrey}, \bibinfo{person}{Marcelo Pasin}, \bibinfo{person}{Pascal Felber}, {and} \bibinfo{person}{Valerio Schiavoni}.} \bibinfo{year}{2022}\natexlab{}.
\newblock \showarticletitle{WebAssembly as a Common Layer for the Cloud-edge Continuum}. In \bibinfo{booktitle}{\emph{Proceedings of the 2nd Workshop on Flexible Resource and Application Management on the Edge}} (Minneapolis, MN, USA) \emph{(\bibinfo{series}{FRAME '22})}. \bibinfo{publisher}{Association for Computing Machinery}, \bibinfo{address}{New York, NY, USA}, \bibinfo{pages}{3–8}.
\newblock
\showISBNx{9781450393102}
\urldef\tempurl%
\url{https://doi.org/10.1145/3526059.3533618}
\showDOI{\tempurl}


\bibitem[Nastic(2024)]%
        {2024selfprovisioningInfrastructure}
\bibfield{author}{\bibinfo{person}{Stefan Nastic}.} \bibinfo{year}{2024}\natexlab{}.
\newblock \showarticletitle{Self-Provisioning Infrastructures for the Next Generation Serverless Computing}.
\newblock \bibinfo{journal}{\emph{SN Computer Science}} \bibinfo{volume}{5}, \bibinfo{number}{6} (\bibinfo{year}{2024}), \bibinfo{pages}{678 -- 693}.
\newblock
\urldef\tempurl%
\url{https://doi.org/10.1007/s42979-024-03022-w}
\showDOI{\tempurl}


\bibitem[Nastic et~al\mbox{.}(2020)]%
        {nastic2020sloc}
\bibfield{author}{\bibinfo{person}{Stefan Nastic}, \bibinfo{person}{Andrea Morichetta}, \bibinfo{person}{Thomas Pusztai}, \bibinfo{person}{Schahram Dustdar}, \bibinfo{person}{Xiaoning Ding}, \bibinfo{person}{Deepak Vij}, {and} \bibinfo{person}{Ying Xiong}.} \bibinfo{year}{2020}\natexlab{}.
\newblock \showarticletitle{Sloc: Service level objectives for next generation cloud computing}.
\newblock \bibinfo{journal}{\emph{IEEE Internet Computing}} \bibinfo{volume}{24}, \bibinfo{number}{3} (\bibinfo{year}{2020}), \bibinfo{pages}{39--50}.
\newblock


\bibitem[Nastic et~al\mbox{.}(2022)]%
        {scf}
\bibfield{author}{\bibinfo{person}{Stefan Nastic}, \bibinfo{person}{Philipp Raith}, \bibinfo{person}{Alireza Furutanpey}, \bibinfo{person}{Thomas Pusztai}, {and} \bibinfo{person}{Schahram Dustdar}.} \bibinfo{year}{2022}\natexlab{}.
\newblock \showarticletitle{A Serverless Computing Fabric for Edge \& Cloud}. In \bibinfo{booktitle}{\emph{2022 IEEE 4th International Conference on Cognitive Machine Intelligence (CogMI)}}. \bibinfo{pages}{1--12}.
\newblock
\urldef\tempurl%
\url{https://doi.org/10.1109/CogMI56440.2022.00011}
\showDOI{\tempurl}


\bibitem[Pusztai et~al\mbox{.}(2022)]%
        {pusztai2022polaris}
\bibfield{author}{\bibinfo{person}{Thomas Pusztai}, \bibinfo{person}{Stefan Nastic}, \bibinfo{person}{Andrea Morichetta}, {et~al\mbox{.}}} \bibinfo{year}{2022}\natexlab{}.
\newblock \showarticletitle{Polaris scheduler: SLO-and topology-aware microservices scheduling at the edge}. In \bibinfo{booktitle}{\emph{2022 IEEE/ACM 15th International Conference on Utility and Cloud Computing (UCC)}}. IEEE, \bibinfo{pages}{61--70}.
\newblock


\bibitem[Rajput et~al\mbox{.}(2022)]%
        {EdgeFaaSBench}
\bibfield{author}{\bibinfo{person}{Kaustubh~Rajendra Rajput}, \bibinfo{person}{Chinmay~Dilip Kulkarni}, {et~al\mbox{.}}} \bibinfo{year}{2022}\natexlab{}.
\newblock \showarticletitle{EdgeFaaSBench: Benchmarking Edge Devices Using Serverless Computing}. In \bibinfo{booktitle}{\emph{2022 IEEE International Conference on Edge Computing and Communications (EDGE)}}. \bibinfo{pages}{93--103}.
\newblock
\urldef\tempurl%
\url{https://doi.org/10.1109/EDGE55608.2022.00024}
\showDOI{\tempurl}


\bibitem[Schleier-Smith et~al\mbox{.}(2021)]%
        {whatServerlessIsAndWhatShouldBecome}
\bibfield{author}{\bibinfo{person}{Johann Schleier-Smith}, \bibinfo{person}{Vikram Sreekanti}, {et~al\mbox{.}}} \bibinfo{year}{2021}\natexlab{}.
\newblock \showarticletitle{What Serverless Computing is and Should Become: The next Phase of Cloud Computing}.
\newblock \bibinfo{journal}{\emph{Commun. ACM}} \bibinfo{volume}{64}, \bibinfo{number}{5} (\bibinfo{date}{apr} \bibinfo{year}{2021}), \bibinfo{pages}{76–84}.
\newblock
\showISSN{0001-0782}
\urldef\tempurl%
\url{https://doi.org/10.1145/3406011}
\showDOI{\tempurl}


\bibitem[Schmid et~al\mbox{.}(2025)]%
        {sebs-flow}
\bibfield{author}{\bibinfo{person}{Larissa Schmid}, \bibinfo{person}{Marcin Copik}, {et~al\mbox{.}}} \bibinfo{year}{2025}\natexlab{}.
\newblock \showarticletitle{SeBS-Flow: Benchmarking Serverless Cloud Function Workflows}. In \bibinfo{booktitle}{\emph{Proceedings of the Twentieth European Conference on Computer Systems}} (Rotterdam, Netherlands) \emph{(\bibinfo{series}{EuroSys '25})}. \bibinfo{pages}{902–920}.
\newblock
\showISBNx{9798400711961}
\urldef\tempurl%
\url{https://doi.org/10.1145/3689031.3717465}
\showDOI{\tempurl}


\bibitem[Shillaker and Pietzuch(2020)]%
        {faasm}
\bibfield{author}{\bibinfo{person}{Simon Shillaker} {and} \bibinfo{person}{Peter Pietzuch}.} \bibinfo{year}{2020}\natexlab{}.
\newblock \showarticletitle{Faasm: Lightweight Isolation for Efficient Stateful Serverless Computing}. In \bibinfo{booktitle}{\emph{2020 USENIX Annual Technical Conference (USENIX ATC 20)}}. \bibinfo{publisher}{USENIX Association}, \bibinfo{pages}{419--433}.
\newblock
\showISBNx{978-1-939133-14-4}


\bibitem[Wang et~al\mbox{.}(2018)]%
        {wang2018peeking}
\bibfield{author}{\bibinfo{person}{Liang Wang}, \bibinfo{person}{Mengyuan Li}, \bibinfo{person}{Yinqian Zhang}, \bibinfo{person}{Thomas Ristenpart}, {and} \bibinfo{person}{Michael Swift}.} \bibinfo{year}{2018}\natexlab{}.
\newblock \showarticletitle{Peeking behind the curtains of serverless platforms}. In \bibinfo{booktitle}{\emph{2018 USENIX Annual Technical Conference (USENIX ATC 18)}}. \bibinfo{pages}{133--146}.
\newblock


\bibitem[Wang(2022)]%
        {wang_how_far_2022}
\bibfield{author}{\bibinfo{person}{Wenwen Wang}.} \bibinfo{year}{2022}\natexlab{}.
\newblock \showarticletitle{How Far We’ve Come – A Characterization Study of Standalone WebAssembly Runtimes}. In \bibinfo{booktitle}{\emph{2022 IEEE International Symposium on Workload Characterization (IISWC)}}. \bibinfo{pages}{228--241}.
\newblock
\urldef\tempurl%
\url{https://doi.org/10.1109/IISWC55918.2022.00028}
\showDOI{\tempurl}


\bibitem[Wen et~al\mbox{.}(2025)]%
        {wen2025unveiling}
\bibfield{author}{\bibinfo{person}{Jinfeng Wen}, \bibinfo{person}{Zhenpeng Chen}, \bibinfo{person}{Federica Sarro}, {and} \bibinfo{person}{Shangguang Wang}.} \bibinfo{year}{2025}\natexlab{}.
\newblock \showarticletitle{Unveiling overlooked performance variance in serverless computing}.
\newblock \bibinfo{journal}{\emph{Empirical Software Engineering}} \bibinfo{volume}{30}, \bibinfo{number}{2} (\bibinfo{year}{2025}), \bibinfo{pages}{1--26}.
\newblock


\bibitem[Xavier et~al\mbox{.}(2015)]%
        {xavier2015performance}
\bibfield{author}{\bibinfo{person}{Miguel~G Xavier}, \bibinfo{person}{Israel~C De~Oliveira}, \bibinfo{person}{Fabio~D Rossi}, {et~al\mbox{.}}} \bibinfo{year}{2015}\natexlab{}.
\newblock \showarticletitle{A performance isolation analysis of disk-intensive workloads on container-based clouds}. In \bibinfo{booktitle}{\emph{2015 23rd Euromicro International Conference on Parallel, Distributed, and Network-Based Processing}}. IEEE, \bibinfo{pages}{253--260}.
\newblock


\bibitem[Xie et~al\mbox{.}(2021)]%
        {xie2021evaluation}
\bibfield{author}{\bibinfo{person}{Dong Xie}, \bibinfo{person}{Yang Hu}, {and} \bibinfo{person}{Li Qin}.} \bibinfo{year}{2021}\natexlab{}.
\newblock \showarticletitle{An evaluation of serverless computing on x86 and arm platforms: Performance and design implications}. In \bibinfo{booktitle}{\emph{2021 IEEE 14th International Conference on Cloud Computing (CLOUD)}}. IEEE, \bibinfo{pages}{313--321}.
\newblock


\bibitem[Zhou et~al\mbox{.}(2017)]%
        {zhou2017performance}
\bibfield{author}{\bibinfo{person}{Li Zhou}, \bibinfo{person}{Yifan Zhang}, \bibinfo{person}{Youhuizi Li}, \bibinfo{person}{Na Yun}, {and} \bibinfo{person}{Lifeng Yu}.} \bibinfo{year}{2017}\natexlab{}.
\newblock \showarticletitle{I/O Performance Isolation Analysis and Optimization on Linux Containers.}. In \bibinfo{booktitle}{\emph{SEKE}}. \bibinfo{pages}{61--66}.
\newblock


\end{thebibliography}
\end{document}